\documentclass[aps,pra,showpacs,superscriptaddress,twocolumn]{revtex4}

\usepackage{amsmath,amsfonts,amssymb,graphicx,epsfig,color,times,wasysym,rotating,multirow}

\newcommand{\hH}{\hat{H}}

\newcommand{\ha}{\hat{a}}
\newcommand{\had}{\hat{a}^\dagger}
\newcommand{\hb}{\hat{b}}
\newcommand{\hbd}{\hat{b}^\dagger}

\newcommand{\htr}{\hat{t}}
\newcommand{\htrd}{\hat{t}^\dagger}

\newcommand{\hc}{\hat{c}}
\newcommand{\hcd}{\hat{c}^\dagger}

\newcommand{\tH}{\tilde{H}}

\newcommand{\tJ}{\tilde{J}}
\newcommand{\tB}{\tilde{B}}

\newcommand{\bra}[1]{\langle #1 \vert}
\newcommand{\brav}{\langle 0 \vert}
\newcommand{\ket}[1]{\vert #1 \rangle}
\newcommand{\ketv}{\vert 0 \rangle}

\newcommand{\expv}[1]{\langle #1 \rangle}

\newcommand{\hadj}{\textrm{H.a.}}

\newcommand{\Tr}{\mathrm{Tr}}

\newcommand{\nr}{n_{\rm r}}

\newcommand{\JA}{J_{\rm A}}
\newcommand{\JB}{J_{\rm B}}

\newcommand{\Ja}{J_{\rm a}}
\newcommand{\Jb}{J_{\rm b}}

\newcommand{\na}{n_{\rm a}}
\newcommand{\nb}{n_{\rm b}}
\newcommand{\Ua}{U_{\rm a}}
\newcommand{\Ub}{U_{\rm b}}
\newcommand{\Uab}{U_{\rm ab}}

\newcommand{\tket}[3]{\ket{#1}^{#2}_{#3}}
\newcommand{\qA}{q_{m}}
\newcommand{\qB}{{1-q_{m}}}
\newcommand{\hadA}{\had_{\rm A}}
\newcommand{\hadB}{\had_{\rm B}}
\newcommand{\haA}{\ha_{\rm A}}
\newcommand{\haB}{\ha_{\rm B}}
\newcommand{\ketvA}{\ketv_{\rm A}}
\newcommand{\ketvB}{\ketv_{\rm B}}
\newcommand{\bravA}{\brav_{\rm A}}
\newcommand{\bravB}{\brav_{\rm B}}

\usepackage[pdftex]{hyperref}

\begin{document}

\title{Dynamics and evaporation of defects 
in Mott-insulating clusters of boson pairs}

\author{Dominik Muth}
\email{muth@physik.uni-kl.de}
\affiliation{Fachbereich Physik und Forschungszentrum OPTIMAS, 
Technische Universit\"at Kaiserslautern, D-67663 Kaiserslautern, Germany}
\affiliation{Graduate School Materials Science in Mainz, 
Technische Universit\"at Kaiserslautern, D-67663 Kaiserslautern, Germany}

\author{David Petrosyan}
\affiliation{Fachbereich Physik und Forschungszentrum OPTIMAS, 
Technische Universit\"at Kaiserslautern, D-67663 Kaiserslautern, Germany}
\affiliation{Institute of Electronic Structure and Laser, FORTH, 
GR-71110 Heraklion, Crete, Greece}

\author{Michael Fleischhauer}
\affiliation{Fachbereich Physik und Forschungszentrum OPTIMAS, 
Technische Universit\"at Kaiserslautern, D-67663 Kaiserslautern, Germany}

\begin{abstract}
Repulsively bound pairs of particles in a lattice governed 
by the Bose-Hubbard model can form stable incompressible clusters 
of dimers corresponding to finite-size $n=2$ Mott insulators.
Here we study the dynamics of hole defects in such clusters 
corresponding to unpaired particles which can resonantly tunnel 
out of the cluster into the lattice vacuum. Due to bosonic statistics, 
the unpaired particles have different effective mass inside and 
outside the cluster, and ``evaporation'' of hole defects from the 
cluster boundaries is possible only when their quasi-momenta are 
within a certain transmission range. We show that quasi-thermalization 
of hole defects occurs in the presence of catalyzing particle defects 
which thereby purify the Mott insulating clusters. 
We study the dynamics of one-dimensional system using 
analytical techniques and numerically exact t-DMRG simulations.
We derive an effective strong-interaction model that enables
simulations of the system dynamics for much longer times.
We also discuss a more general case of two bosonic species which 
reduces to the fermionic Hubbard model in the strong interaction limit.
\end{abstract}


\date{\today}
\maketitle


\section{Introduction}

Quantum particles in lattice potentials, e.g. electrons in crystals, have been 
studied since the early days of quantum theory \cite{Ashcroft1976,Fetter1971}. 
With the development of artificial (optical) lattice potentials 
for cold neutral atoms \cite{Bloch2008}, bosonic lattice models 
are recently attracting increased interest as well \cite{Lewenstein2007},
with the Bose-Hubbard model (BHM) \cite{Fisher1989} being an important 
example. A remarkable phenomenon entailed by the BHM is that pairs 
of strongly interacting bosons can form tightly bound ``dimers'' 
both for attractive and repulsive interactions 
\cite{Winkler2006, Piil2007, Valiente2008a, Petrosyan2007}. 
In free space, or in the presence of energy dissipation, the repulsive 
interaction inevitably leads to pair dissociation. In a lattice, 
however, the kinetic energy of each particle is restricted to the 
values in the allowed Bloch band. Consequently, two co-localized
particles in a dissipation-free lattice remain tightly bound together 
as a dimer when their interaction energy $U$ exceeds the kinetic 
energy of free particles $\sim J$ within the Bloch band.

In a previous publication \cite{Petrosyan2007}, we have studied 
the many-body dynamics of the repulsively-bound dimers of bosons. 
Due to virtual transitions of the dimer constituent particles, 
the dimers at the neighboring lattice sites strongly attract each 
other, with the corresponding interaction energy exceeding the 
dimer tunneling energy by a factor of 4. For many dimers on the lattice, 
it is then energetically favorable to form dynamically stable ``droplets'', 
constituting incompressible Mott-insulating (MI) clusters with 
the number of particles per site of exactly $n=2$. 
Inevitable imperfections in the preparation process would typically
cause such MI clusters to contain hole and particle 
defects corresponding, respectively, to unpaired and excess 
particles (monomers and trimers). An important question is thus 
how to purify the system of the defects reducing thereby the entropy. 
In the present paper, we discuss a mechanism of self-purification 
of stable MI clusters of dimers surrounded by lattice vacuum. 
We study the dynamics of defects in one-dimensional system 
by analytical calculations and numerical many-body simulations.

Within the cluster, hole and particle defects can propagate via resonant 
single-particle hopping with enhanced amplitude, which stems from 
the bosonic statistics of the surrounding $n=2$ MI environment.
Outside the cluster, hole defects correspond to free particles. 
Since their tunneling energy $J$ is much larger than the monomer-dimer 
interaction energy $\sim J^2/U$ \cite{Valiente2010a}, hole defects are 
not bound to the cluster and can ``evaporate''. However, the widths 
of the single particle Bloch band is twice larger inside the cluster 
than outside of it, therefore only the hole defects with energies in 
the center of the band can penetrate the cluster boundaries and evaporate
into the lattice vacuum, while in the absence of quasi-momentum redistribution, 
low- and high-energy hole defects will remain in the cluster. 
We show that the presence of particle defects leads to efficient 
``thermalization'' of the hole defects via quasi-momentum redistributing 
collisions. Hence, very few such ``catalyzing''  particle defects 
can purify the MI cluster. 

Before continuing, we note a recent relevant work \cite{Heidrich-Meisner2009}
dealing with fermionic dimers described by the Hubbard model. After preparing 
a cold atomic gas with filling of $n \simeq 2$ in the trap center, followed 
by turning off the trap, the hole defects will simply tunnel out of the cluster
into the vacuum. For fermions, however, the remaining cluster is not stable, 
since the effective second-order tunneling of the on-site pairs is not 
restrained by the interaction between the pairs.  

\begin{figure}[t]
\flushleft{(a)}\\[2pt]
\centering \includegraphics[width=.35\textwidth]{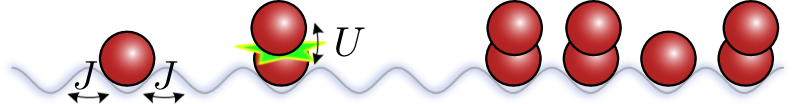}\\
\flushleft (b)\\[2pt]
\centering \includegraphics[width=.35\textwidth]{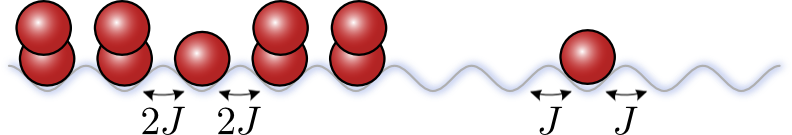}\\
\flushleft (c)\\[2pt]
\centering \includegraphics[width=.35\textwidth]{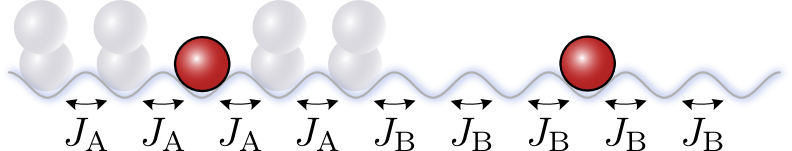}\\
\flushleft (d)\\[1pt]
\centering \includegraphics[width=.35\textwidth]{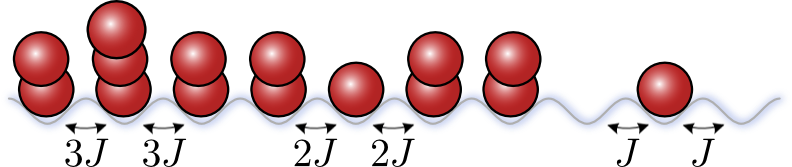}\\
\flushleft (e)\\[2pt]
\centering \includegraphics[width=.35\textwidth]{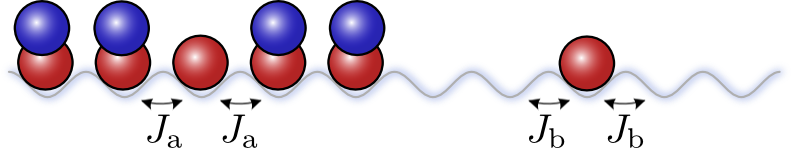}
\caption{(Color online) Physical models studies in this paper.
(a)~The Bose-Hubbard model, Eq.~(\ref{eq:HBH}).
(b)~Monomer (hole defect) effective hopping, Sec.~\ref{sec:single}.
(c)~Single defect effective theory, Sec.~\ref{sec:single}.
(d)~Trimer (particle defect) effective hopping, Sec.~\ref{sec:momrd}.
(e)~Single hole defect in two species Bose-Hubbard model, 
Sec.~\ref{sec:twospecies}.}
\label{fig:particles}
\end{figure}

Figure~\ref{fig:particles} illustrates the main physics studied 
in this paper, which is organized as follows. 
In Sec.~\ref{sec:pairsonly} we outline the properties of the pure dimer 
clusters \cite{Petrosyan2007}. We then introduce in Sec.~\ref{sec:single} 
an effective theory of scattering of a single particle (hole defect) 
from a domain wall separating the dimer cluster and the vacuum. 
The quasi-momentum redistribution of a hole defect upon collisions with 
a particle defect in the lattice with periodic and open boundaries 
is studied in Sec.~\ref{sec:momrd}. In Sec.~\ref{sec:manybodydyn} 
we present the results of many-body numerical simulations for a realistic 
system with several hole and particle defects in a dimer cluster surrounded 
by empty lattice. Finally, in Sec.~\ref{sec:twospecies} we discuss the case 
of two bosonic species, which is more flexible theoretically, but is 
demanding experimentally. In the limit of infinite intra-species interaction,
it contains the special case of the Hubbard model \cite{Heidrich-Meisner2009},
since in one dimension and in the absence of double occupancy, 
bosons and fermions are equivalent through the Jordan-Wigner transformation. 
Much of the involved technical details are deferred to Appendices 
\ref{sec:apptrans}, \ref{sec:appmomentum}, \ref{sec:appeffective} 
and \ref{sec:appconstr}.

\section{Repulsively bound dimers}
\label{sec:pairsonly}

The underlying Hamiltonian for our system is that 
of the BHM \cite{Fisher1989} 
\begin{equation}
\hH = -J \sum_j ( \hbd_j\hb_{j+1}  +\hadj ) 
+ \mbox{$\frac{1}{2}$} U \sum_j \hbd_j \hbd_j \hb_j \hb_j , \label{eq:HBH}
\end{equation}
where $\hb$ and $\hbd$ are bosonic annihilation and creation operators, 
$J$ is the particle hopping rate between adjacent lattice sites $j, j+1$, 
and $U$ is the contact interaction between the particles on the same lattice site. 
Throughout this paper, we assume that the on-site interaction is the dominant 
energy parameter, $U\gg J$.

Considering first a lattice containing only zero or two particles per site, 
we do not allow the dimer occupation number in the lattice to exceed unity. 
Adiabatically eliminating all the states with odd number of particles 
per site, we obtain for the dimers an effective Hamiltonian 
\cite{Petrosyan2007} that contains only terms with characteristic 
energies on the scale of $J^2/U \ll J$:
\begin{equation}
\hH = -\tJ \sum_j (\hcd_j\hc_{j+1}+\hadj ) +
\tB \sum_j \hcd_j\hcd_{j+1}\hc_{j+1}\hc_j, \label{eq:Hpair}
\end{equation}
where $\tJ=-2J^2/U$ is the dimer hopping rate and $\tB=-16J^2/U$
is the nearest neighbor interaction. The dimer creation $\hcd_j$ 
and annihilation $\hc_j$ operators satisfy the hard-core boson 
commutation relations
\begin{subequations}
\label{eq:hardcore}
\begin{eqnarray}
 i\ne j: &\ & [\hc_i, \hc_j] = [ \hc_i, \hcd_j] =0 , \\
 i =  j: &\ & \{ \hc_j , \hc_j \} =0, \, \{ \hc_j, \hcd_j\} = 1 . 
\end{eqnarray}
\end{subequations}
Hamiltonian~(\ref{eq:Hpair}) can be mapped onto that 
for the spin-$\frac{1}{2}$ $XXZ$ model \cite{Schmidt2009a, Muth2010} 
with the anisotropy parameter $\Delta = \tB/2\tJ = 4$. For $\Delta > 1$, 
we are in the ferromagnetic, Ising-like regime, and a cluster of dimers, 
corresponding to a lattice domain with maximum magnetization, 
is dynamically stable. To understand this in terms of dimers, 
observe that, for any $U(\gg J)$, the maximal kinetic energy $2\tJ$ 
gained by releasing a dimer from the cluster boundary is small 
compared to the binding energy $\tB$ of the dimer to the cluster.

The stability of the dimer cluster is an intrinsic feature of the BHM. 
It is rooted in the bosonic amplification of the inter-site hopping 
of the particles, which in turn enhances the effective (second-order) 
nearest neighbor interaction $\tB$. For the fermionic Hubbard model 
discussed in \cite{Heidrich-Meisner2009} in the context 
of defect evaporation from a dimer cluster, i.e., a band insulator, 
we show in Sec.~\ref{sec:twospecies} that $\Delta = 1$ ($\tB = 2\tJ$), 
which means that the cluster is unstable and the dimers will diffuse away.

\section{Single defect model in the strong-interaction limit}
\label{sec:single}

The dynamics of dimers is rather slow, as it is governed by 
the small characteristic energies $\sim J^2/U$, but the dynamics 
of monomers is much faster, involving large single-particle hopping 
rate $J$. We can thus retain only the contributions on the scale 
of $J$, which results in a very simple and transparent effective 
theory for the monomers.
For a monomer in the cluster (hole defect), the bosonic statistics 
plays an important role: it increases the hopping amplitude of the
monomer in the environment of dimers by a factor of $2$,
see Fig.~\ref{fig:particles}(b). As a result, the kinetic energy 
of the monomer in the dimer cluster is $E_k = -4J\cos(k)$, 
while in the vacuum it is $E_k = -2J\cos(k)$, where $k \in [-\pi, \pi]$ 
is the monomer quasi-momentum quantified by the phase change between 
neighboring lattice sites. Therefore the monomer will be confined 
to the cluster if its quasi-momentum is not inside the transmission region 
given by
\begin{equation}
\label{eq:allowedk}
 k \in (-2\pi/3,-\pi/3) \cup (\pi/3,2\pi/3),
\end{equation}
up to a correction due to small interactions of the order of $J^2/U \ll J$ 
which we neglected.

Consider the scattering of a monomer from the domain wall between 
the dimer cluster occupying sites $j < 0$ (region A) and the vacuum 
at sites $j > 0$ (region B), see Fig.~\ref{fig:particles}(c). 
The local bare particle number is $n_j=2$ for $j<0$ and $n_j=0$ for $j>0$. 
The particle number at $j=0$ depends on the position $i$ of the monomer
$n_i=1$: inside the cluster $i<0$ we have $n_0=2$, at the boundary $i=0$ 
obviously $n_0=1$, while outside the cluster (in the vacuum) $i>0$ leads 
to $n_0=0$. Hence the position of the wall shifts upon the monomer 
crossing the boundary, which should be taken into account when considering 
many defects. The hopping rate of the monomer is $\JA$ for sites $j \leq 0$ 
and $\JB$ for sites $j > 0$. The effective Hamiltonian for a single 
monomer then reads 
\begin{equation}
\hH = -\JA \sum_{j < 0} (\had_j\ha_{j+1}  + \hadj) 
- \JB \sum_{j \ge 0} (\had_j\ha_{j+1} + \hadj) , \label{eq:HTrans}
\end{equation}
with $\JA = 2J$ and $\JB = J$.

\begin{figure}[t]
\includegraphics[width=.6\columnwidth]{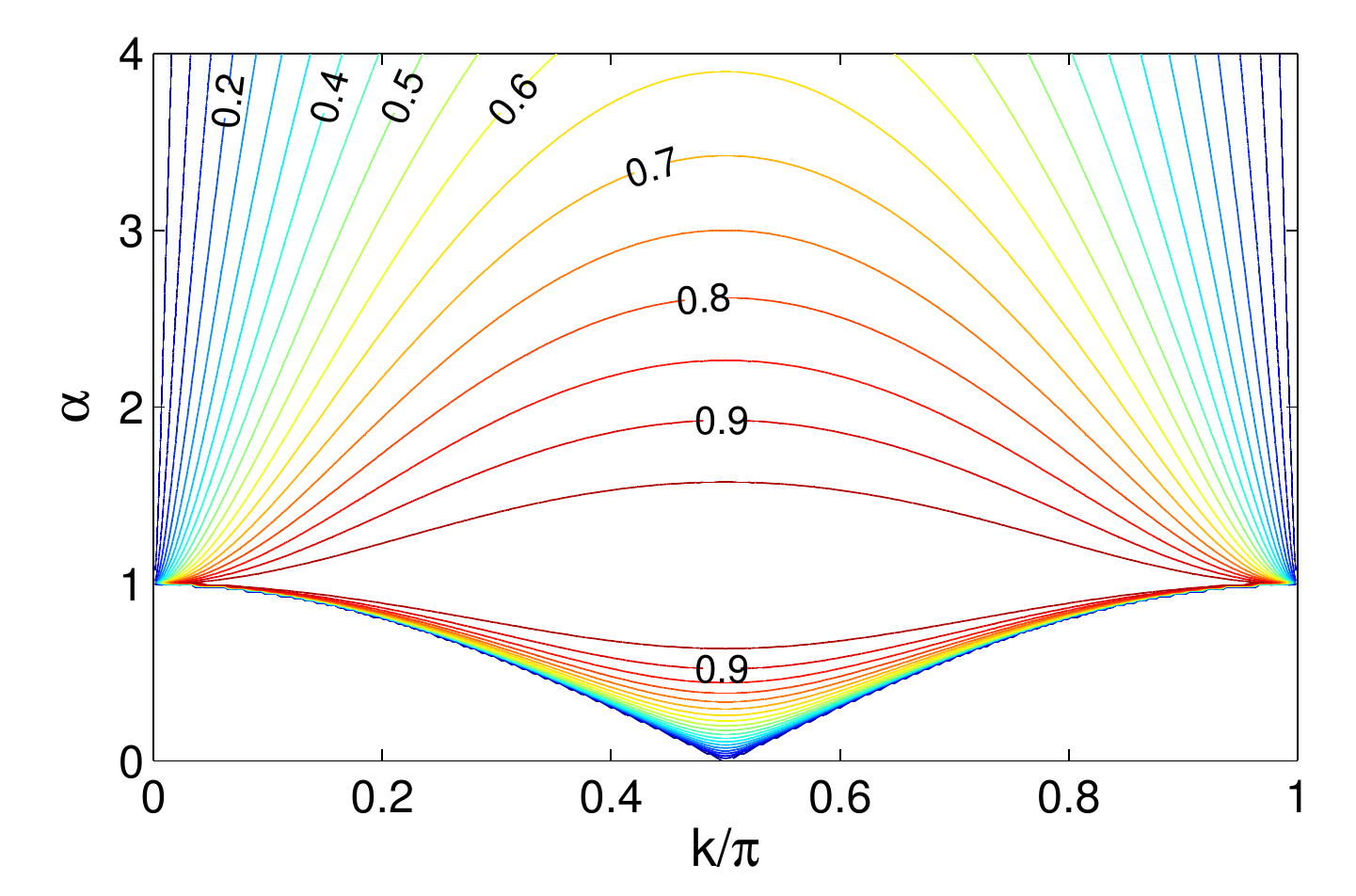}\\
\includegraphics[width=.5\columnwidth]{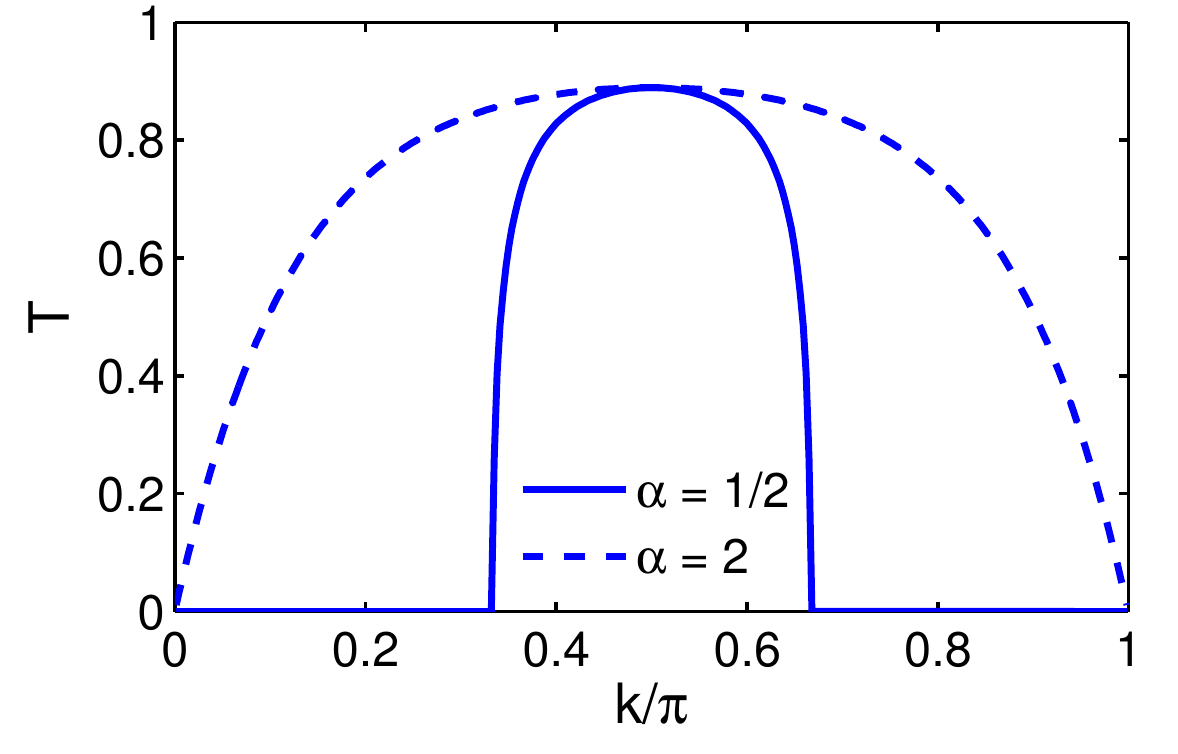}%
\includegraphics[width=.5\columnwidth]{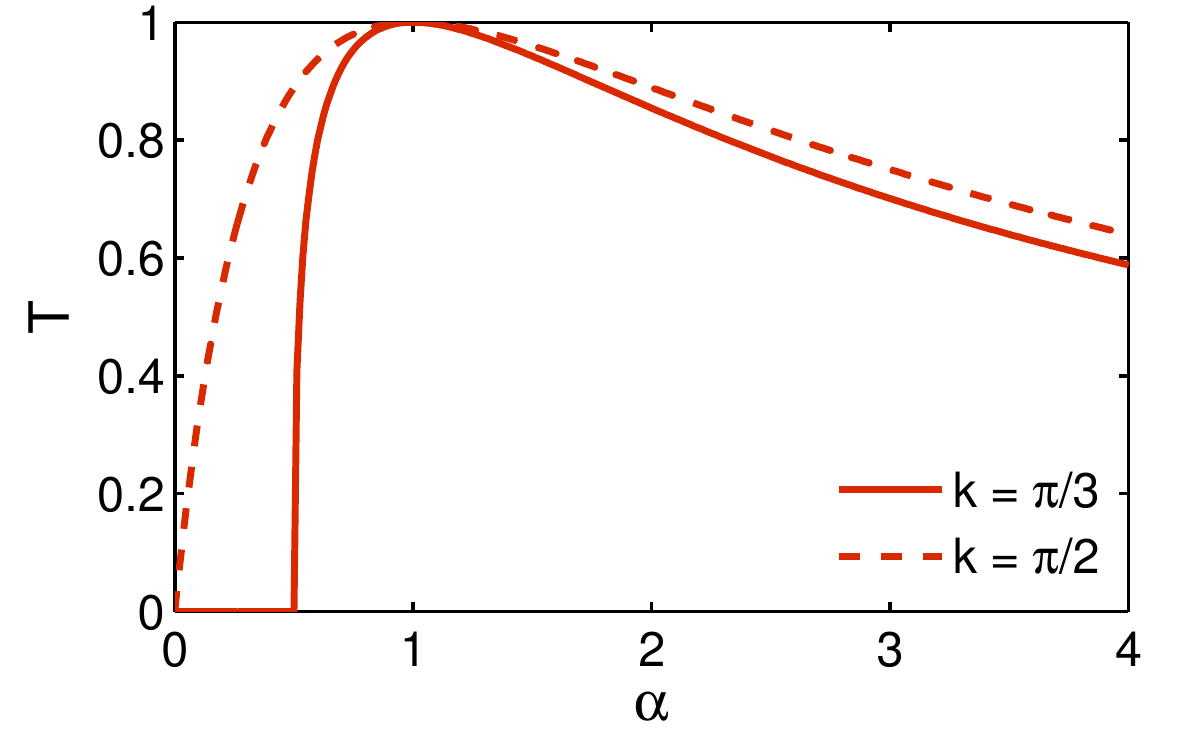}%
\caption{(Color online) Transmission probability $T(k)$ 
[cf. Eq.~(\ref{eq:transm})] for various $\alpha = J_{B}/J_{A}$.}
\label{fig:Talphak}
\end{figure}

In Appendix \ref{sec:apptrans} we calculate the exact transmission 
probability $T(k)$ of a particle crossing a domain wall in a system 
described by Hamiltonian (\ref{eq:HTrans}). The results are illustrated 
in Fig. \ref{fig:Talphak} for various $\alpha = J_{B}/J_{A}$. 
The values of $\alpha=1/2$ and $\alpha=2$ correspond, respectively, to 
the single particle leaving the dimer cluster and entering it from vacuum.

\section{Quasi-momentum redistribution of the defects}
\label{sec:momrd}

We have seen above that a hole defect can leave the MI cluster only if 
its quasi-momentum is within the transmission range of Eq.~(\ref{eq:allowedk}), 
while a defect with the quasi-momentum outside the transmission range will remain 
in the cluster indefinitely. Hence, to completely purify the cluster of 
hole defects, their quasi-momenta should be continuously redistributed over the entire 
range of $k \in [-\pi,\pi]$. In two or more dimensions, collisions between 
identical particles can redistribute the absolute values of their quasi-momenta, 
and we therefore expect the evaporation of all the defects through the
cluster boundaries after a few collisions. 
In a one dimensional lattice, however, collisions of two particles interacting 
via any finite range potential can only exchange their quasi-momenta or leave 
the quasi-momenta unchanged \cite{Piil2007, Valiente2008a, Valiente2009, Valiente2010b}
(for indistinguishable particles both outcomes are equivalent), provided that the
lattice is deep enough so that the large band gap precludes interband transitions. 
Similarly, collisions with the fixed boundaries can only reverse the quasi-momentum 
of a particle. The simplest quasi-momentum redistribution mechanism is then 
three particle collisions. This happens at a rate proportional to the defect 
density squared, which is too slow for cold atom experiments.

In the dimer cluster, in addition to the hole defects (monomers), 
we may have particle defects (trimers) with different effective mass. 
The hopping rates of a monomer and a trimer in the cluster are 
$J_a = 2J$ and $J_t = 3J$, respectively, Fig. \ref{fig:particles}(d). 
Before collision, their quasi-momenta are $k_a$ and $k_t$, while conservations 
of quasi-momentum, $k_a+k_t=k_a'+k_t'$, and energy, 
$J_a \cos(k_a) + J_t \cos(k_t) = J_a \cos(k_a') + J_t \cos(k_t')$, during 
the collision determine the new quasi-momenta $k_a'$ and $k_t'$ via
\begin{equation}
\label{eq:j1j2scatter}
 J_a\cos(k_a) + J_t \cos(k_t) = J_a \cos(k_a') + J_t \cos(k_a+k_t-k_a').
\end{equation}
If there is a collision with the wall, or a third defect of either kind, before 
this process is reversed, all energetically allowed combinations of $k_a,k_t$ 
can be assumed, as will be verified below by exact numerical simulations.

\subsection{Two classical particles}

The timescale for quasi-momentum redistribution can be calculated from purely 
classical considerations. A monomer or a trimer moving in the MI cluster 
has a kinetic energy of $E_{k_{\mu}} = -2J_{\mu}\cos(k_{\mu})$ and the 
corresponding group velocity of $v_{\mu} = 2J_{\mu} \sin (k_{\mu})$ [$\mu =a,t$].

Consider first two wave packets in a periodic lattice of length $L$. 
After a collision (the defects can not penetrate each other), their 
velocities are assumed to be $v_a = 2J_a\sin (k_a) < v_t = 2J_t\sin (k_t)$. 
The next collision happens after time
\begin{equation}
\frac{t_{\mathrm{c}}}{2}
= \frac{L-1}{2}\frac{1}{J_t\sin (k_t)-J_a\sin (k_a)} ,
\end{equation}
and the new quasi-momenta are determined by Eq.~(\ref{eq:j1j2scatter}). 
It follows that the time interval between all subsequent collisions 
is the same $t_{\mathrm{c}}/2$, since Eq.~(\ref{eq:j1j2scatter}) and 
\begin{eqnarray*}
& & \frac{L-1}{2}\frac{1}{J_t\sin (k_t)-J_a\sin (k_a)}  \\ 
&=& \frac{L-1}{2}\frac{1}{-J_t\sin(k_a+k_t-k_a') +J_a\sin (k_a')}
\end{eqnarray*}
always have a common solution.

In the presence of a wall, or a third defect, the quasi-momenta can take any values 
energetically allowed. A revival is not expected, but now $t_{\mathrm{c}}^{-1}$
is an effective rate of quasi-momentum redistribution. It is essentially given by 
$J$ over the mean free path, i.e., it is proportional to $J$ times the average 
defect density, which is indeed much faster than the rate of three particle 
collisions.

\subsection{Two quantum particles: Numerical simulations}
\label{sec:momrd:ed}

\begin{figure}[t]
\centering
\includegraphics[width=.5\columnwidth]{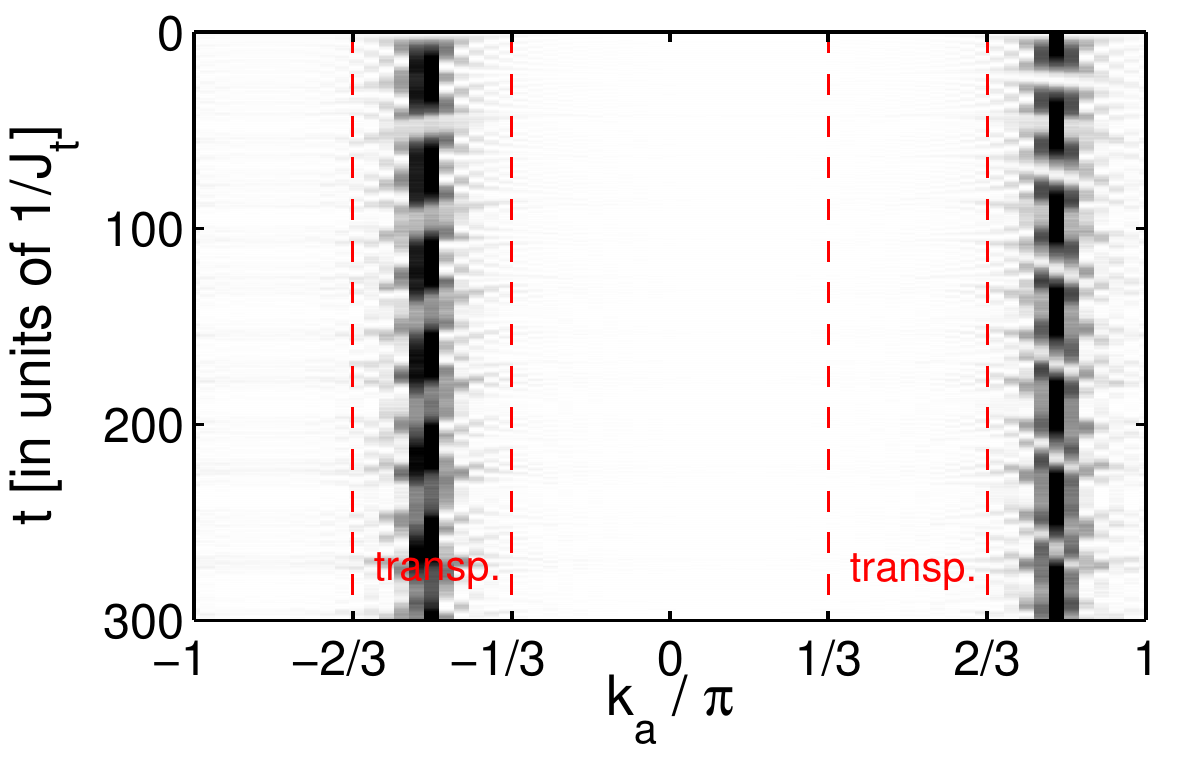}%
\includegraphics[width=.5\columnwidth]{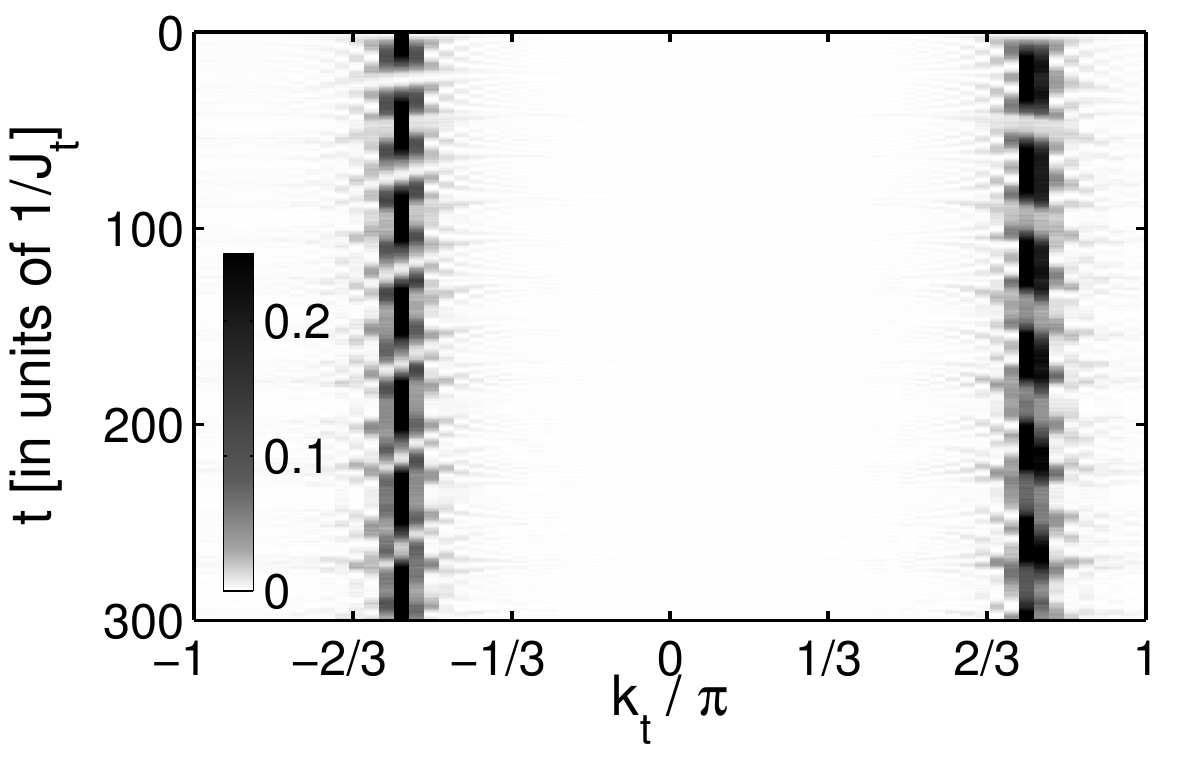}\\
\includegraphics[width=.5\columnwidth]{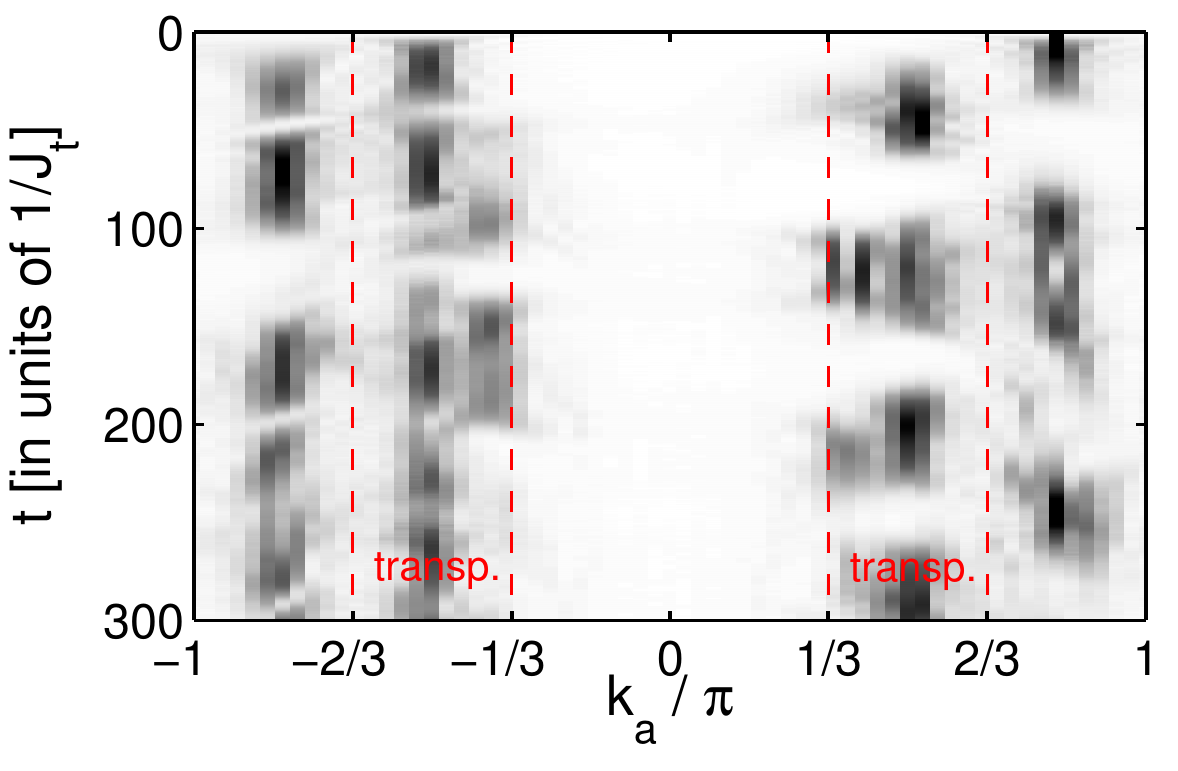}%
\includegraphics[width=.5\columnwidth]{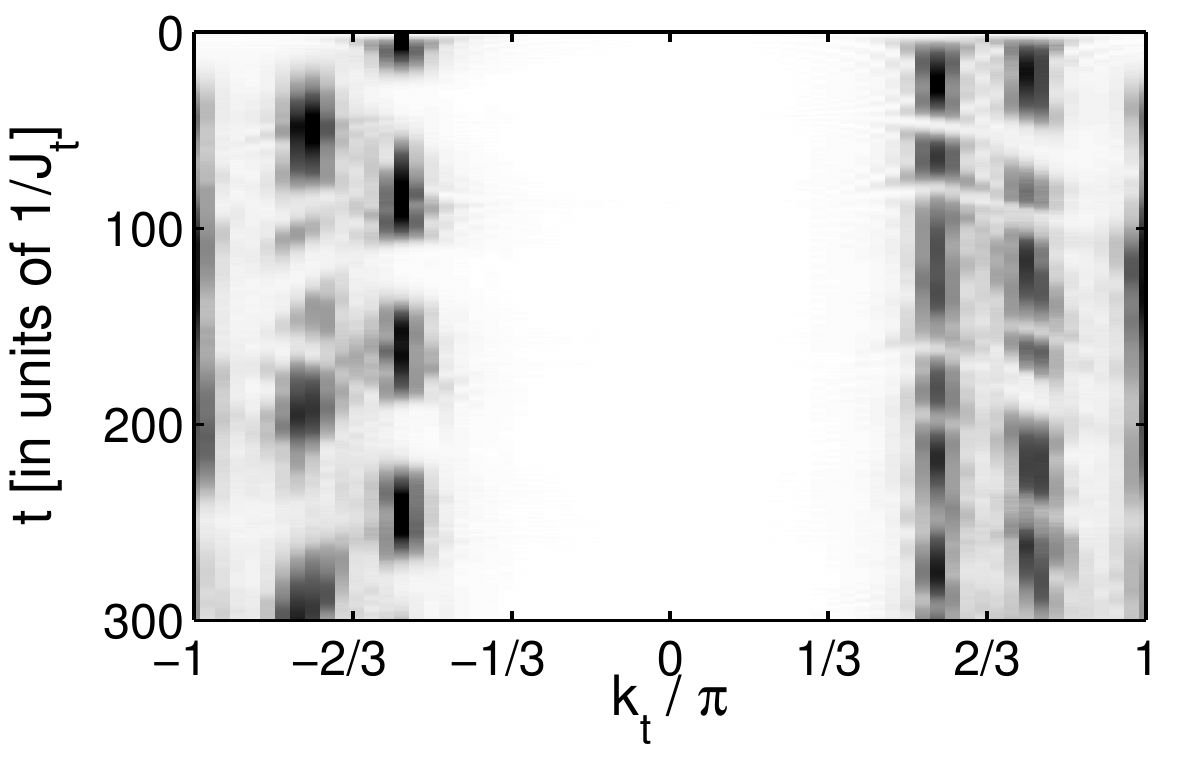}%
\caption{(Color online) Dynamics of the quasi-momentum distribution for the 
monomer (left column) and trimer (right column) in a lattice of 
$L=64$ sites. The initial quasi-momenta are $k_a = \frac{13}{16} \pi$ and 
$k_t = - \frac{9}{16} \pi$. 
Upper panels correspond to periodic boundary conditions, where the 
markers on the right indicate multiples of the revival time 
$t_{\rm c}\approx46.63/J_t$. 
Lower Panels are obtained for open boundary conditions.
Dashed vertical lines mark the transmission regions for 
monomer quasi-momenta as per Eq.~(\ref{eq:allowedk}).}
\label{fig:twodistinct}
\end{figure}

We simulate the quantum dynamics of the hole and particle defects
in a dimer cluster using the two-particle Hamiltonian in quasi-momentum space, 
see Appendix \ref{sec:appmomentum}. Each defect is initially 
prepared in a quasi-momentum eigenstate, with the combined state given by
\begin{equation}
\label{eq:twoinit}
 \ket{k_a, k_t} = \frac{1}{L}\sum_{j_a,j_t=1}^{L}
 e^{ik_aj_a}e^{ik_tj_t}
\had_{j_a}\htrd_{j_t} \ket{\rm vac},
\end{equation}
where $\had_{j}$ creates a monomer and $\htrd_{j}$ a trimer at site $j$ 
of a finite lattice filled with dimers, playing here the role of an effective
vacuum $\ket{\rm vac}$. 

Figure \ref{fig:twodistinct} shows the results of our numerical simulations.
In the case of periodic boundary conditions, the dynamics is mainly classical;
only two values of quasi-momentum $k$ are assumed by each particle, and after the 
classical revival time $t_{\rm c}$, the quasi-momentum distribution is restored to 
the initial. For open boundary conditions, however, we observe fast 
redistribution of quasi-momenta, and already the first revival is hardly noticeable.
We may therefore conclude that a single trimer can catalyze the redistribution 
of quasi-momenta of monomers, making the evaporation of almost all hole defects 
possible, provided that their average kinetic energy is initially close 
to the center of the band. This will be verified by the following many-body 
calculations.

\section{Many-body numerical simulations}
\label{sec:manybodydyn}

To study the dynamics of several defects under experimentally realistic 
conditions, we use a sufficiently long lattice that can accommodate dimer 
clusters spanning a few dozen sites. The complete Hilbert space for such 
a system is too large to be amenable to exact diagonalization treatments. 
We therefore resort to time-dependent density matrix renormalization 
group (t-DMRG) methods \cite{Daley2004,White2004}, specifically, 
the time evolving block decimation (TEBD) algorithm \cite{Vidal2003} 
using the matrix product state (MPS) formalism. 
Even then, however, simulating the full BHM is a difficult task. 
This is due to sizable quantum fluctuations present even in 
the pure dimer cluster for any finite interaction strengths $U/J$. 
These fluctuations contribute to the many-body entanglement and 
consume much of the computational resources required to simulate 
the dynamics of the defects. We therefore introduce an effective 
model for the defects only.

\subsection{Many defect effective theory in the strong-interaction limit}
\label{sec:effdefHam}

Since the states with different number of particles per site have energies 
separated by multiples of $U$($\gg J$), the numbers of monomers, dimers, 
and trimers in a lattice are, to a good approximation, conserved separately. 
This allows us to treat the monomers, dimers, and trimers as distinguishable
species, each represented by hard-core bosons, Eqs.~(\ref{eq:hardcore}). 
Furthermore, as discussed in Sec.~\ref{sec:pairsonly}, dimers forming 
stable clusters do not contribute to the dynamics of the system. 
For our initial conditions, typically containing a single cluster, 
we can thus reformulate the problem as one of the hole and particle 
defects moving on the background of dimers or vacuum, with the spatial 
configuration of the dimer cluster entering the effective Hamiltonian 
for the defects only as a parameter.

We define the reference system in which the pure dimer cluster occupies 
certain lattice sites while all the defects are placed at the beginning
(left side) of the lattice. As the defects move in the lattice, the 
effective hopping rates depend on whether they are inside or outside 
the MI cluster. In turn, the position of the cluster depends on the 
positions of the defects, since each defect crossing the system from 
the left to the right shifts the position of the dimers, and the cluster 
as a whole, by one site to the left. The effective Hamiltonian for the 
defects can then be cast as
\begin{equation}
 \hH = \sum_{j=1}^{L-1} \sum_{\nr = 0}^N \hH_j^{[\Theta(j+\nr)]} 
   \otimes \hat{P}_{[j+2, L]}^{\nr} \equiv \sum_{j=1}^{L-1}\tH_j , 
\label{eq:Heff}
\end{equation}
where $\hat{P}_{[j+2, L]}^{\nr}$ is the projector onto the subspace 
containing exactly $\nr$ hole and particle defects on sites $j+2$ to $L$, 
while each local operator $\hH_j^{[\Theta]}$ acts on sites $j$ and $j+1$ as
\begin{eqnarray}
 \hH_j^{[\Theta]} &=&
 - J_{\rm a}^{[\Theta]} (\had_j\ha_{j+1} + \hadj) 
\htr_j\htr_{j+1}\htrd_{j+1}\htrd_j
\nonumber \\ & & 
 - J_{\rm t}^{[\Theta]} ( \htrd_j\htr_{j+1} + \hadj ) 
\ha_j\ha_{j+1}\had_{j+1}\had_j .
\label{eq:extHubtrim}
\end{eqnarray}
Here $\had_j$ and $\ha_j$ ($\htrd_j$ and $\htr_j$) are the hard-core 
bosonic creation and annihilation operators for the monomers (trimers).  
The function $\Theta(j)$ is initialized for all $j$ with respect to the 
reference system, and it can take two values: $\Theta(j)=1$ for site 
$j+1$ being empty (vacuum) and $\Theta(j)=2$ for site $j+1$ containing 
a dimer. Then the hopping rates for the monomers are 
$J_{\rm a}^{[1]} = J$ and $J_{\rm a}^{[2]} = 2J$, and for the trimers
are $J_{\rm t}^{[1]} = 0$ (they can not move on an empty lattice in
first order in $J$) and $J_{\rm t}^{[2]} = 3J$.

Note that since the effective Hamiltonian (\ref{eq:extHubtrim}) contains
two species of particles with hardcore interactions, it can not be mapped 
onto a model of free fermions via the Jordan-Wigner transformation (which
is possible for identical hardcore bosons). The dynamics is therefore 
non-trivial and actual calculations again require numerical many-body 
(TEBD) techniques. The practical advantage of the effective model---besides
the largely reduced number of particles---is that the fast timescale $U^{-1}$
is eliminated from the system's dynamics and in our numerical simulations 
we can choose Trotter steps on the time scale $\lesssim J^{-1}$. Further 
discussion on the effective defect model is given in 
Appendix~\ref{sec:appeffective}. 

\subsection{Initial states}
\label{sec:MBinst}

In our numerical calculations, we use several typical configurations 
of the defects in the lattice, each configuration described by a pure 
quantum state. Various coherent and incoherent superpositions of such 
configurations would represent mixed initial states. 

We consider piecewise product states. A MI segment of length $l$ contains
fixed number of particles $n$ at every site ($n=2$ inside the dimer cluster 
and $n=0$ in the vacuum),
\begin{equation}
 \tket{\cdot}{n}{l} 
 = \bigotimes_{j=1}^l \frac{(\had_j)^n}{\sqrt{n!}}\ket{\rm vac},
\end{equation}
with $\ket{\rm vac}$ denoting the true vacuum. Each segment can contain 
an additional defect. For a defect localized as site $j$, we use the notation
\begin{subequations}
\begin{eqnarray}
\tket{j_+}{n}{l} &=& \frac{\had_j}{\sqrt{n+1}} \tket{\cdot}{n}{l} , \\
\tket{j_-}{n}{l} &=& \frac{\ha_j}{\sqrt{n}} \tket{\cdot}{n}{l}  
\quad (n \geq 1) ,
\end{eqnarray}
\end{subequations}
with $\pm$ corresponding, respectively, to a particle and a hole defect. 
Similarly, we denote a defect with quasi-momentum $k$, which must be a multiple of 
$2\pi/l$, as
\begin{subequations}
\begin{eqnarray}
\tket{k}{n+}{l} &=& \frac{1}{\sqrt{l}}\sum_{j=1}^{l} e^{ikj} 
\tket{j_+}{n}{l} , \\
\tket{k}{n-}{l} &=& \frac{1}{\sqrt{l}}\sum_{j=1}^{l} e^{ikj} 
\tket{j_-}{n}{l}  \quad (n \geq 1) .
\end{eqnarray}
\end{subequations}

We prepare the cluster by joining MI segments with and without defects.
Since we are only interested in low defect densities, we do not construct 
segments containing multiple defects. In order to perform TEBD simulations, 
the initial states have to be represented in the MPS form, which is 
discussed in Appendix \ref{sec:appconstr}.

\subsection{Numerical results}
\label{sec:numresult}

\begin{figure}[t]
\centering
\includegraphics[width=.5\columnwidth]{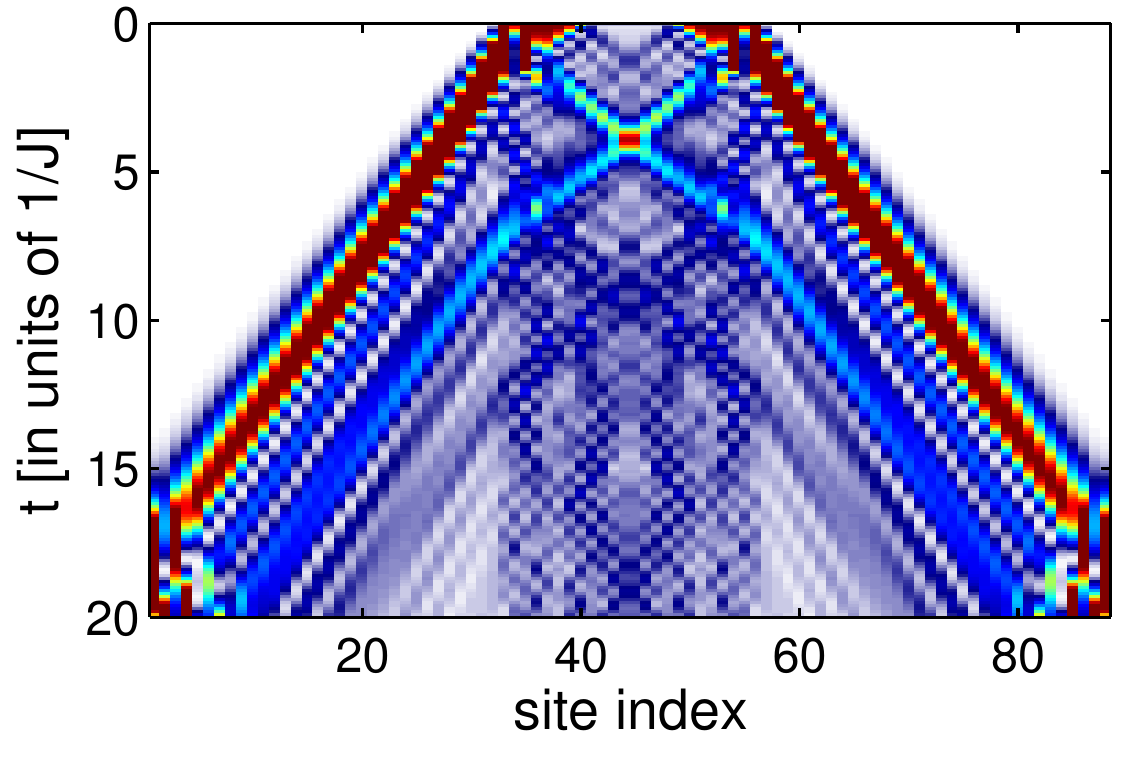}%
\includegraphics[width=.5\columnwidth]{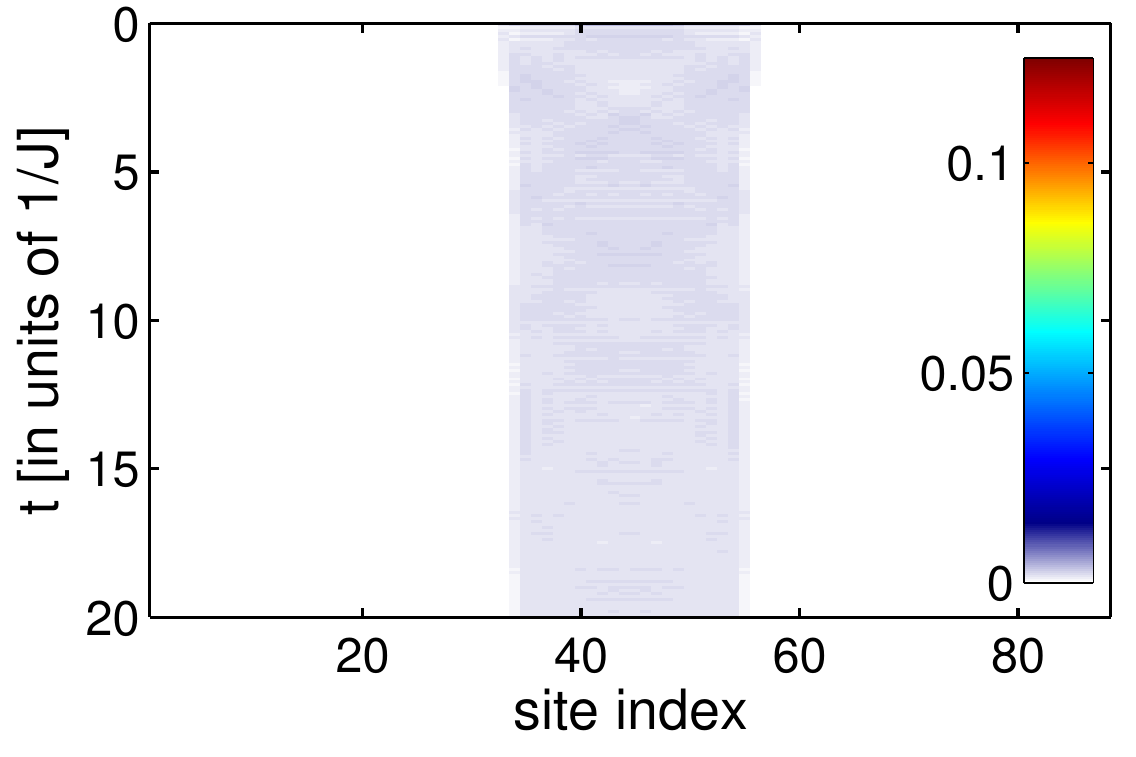}\\
\includegraphics[width=.5\columnwidth]{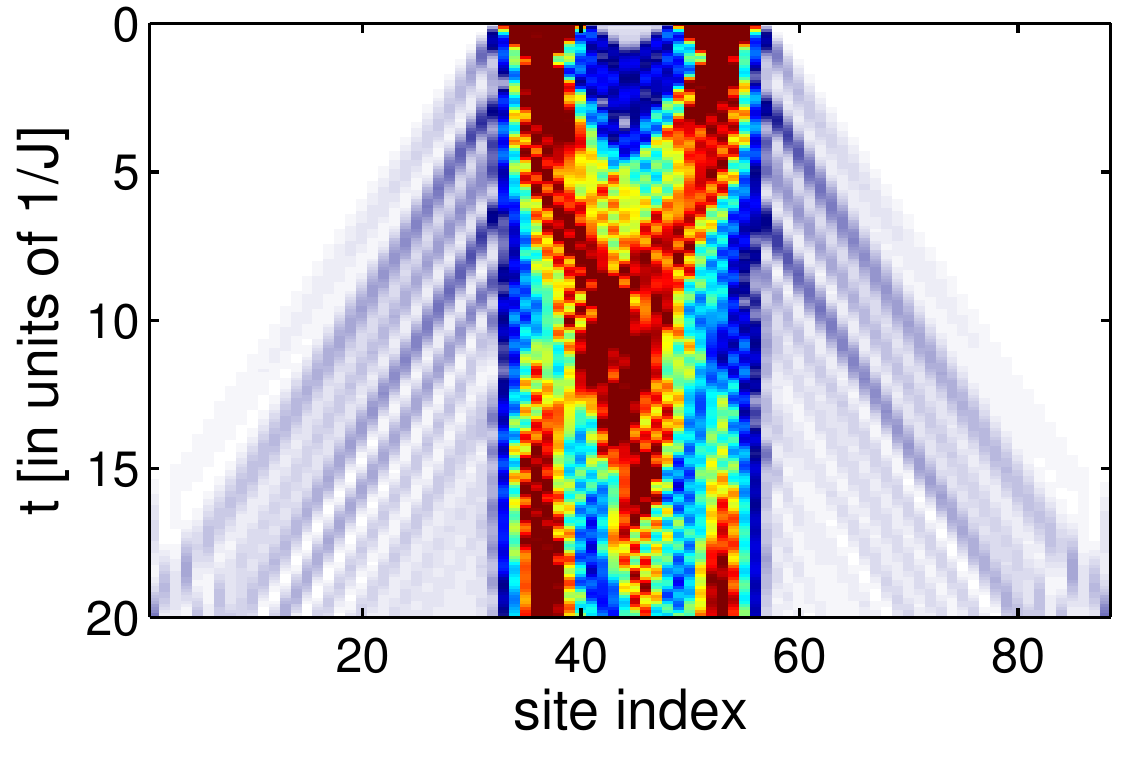}%
\includegraphics[width=.5\columnwidth]{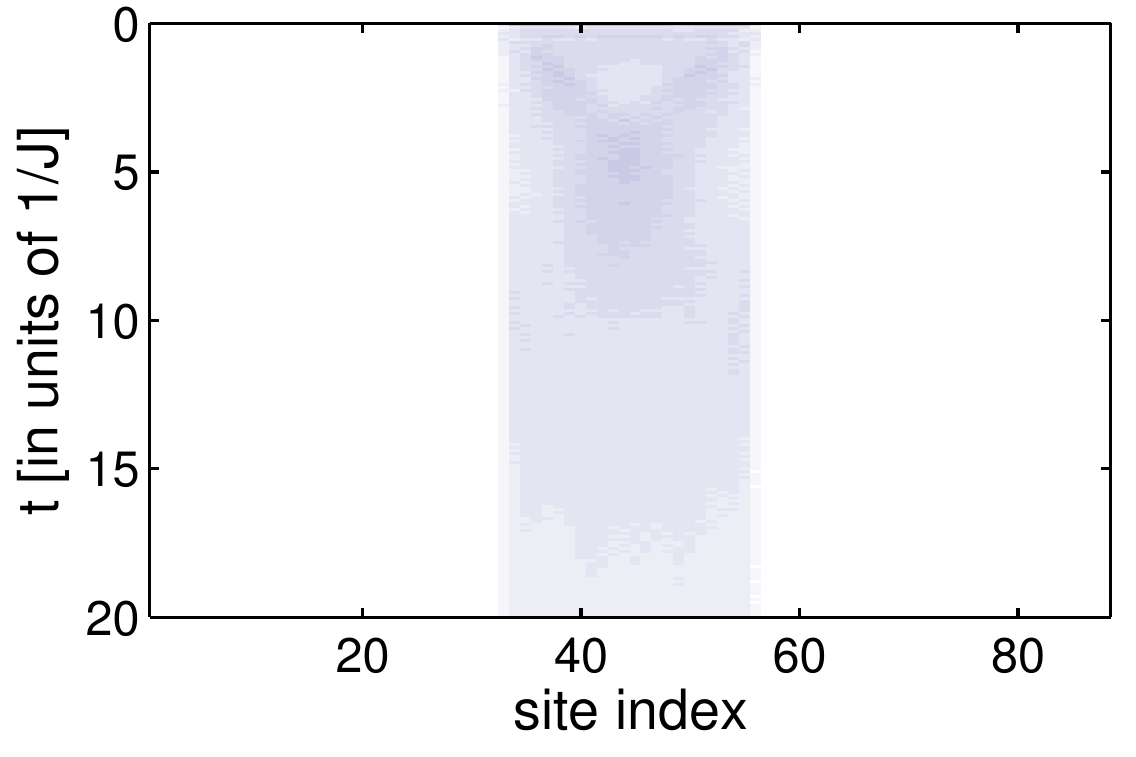}\\
\includegraphics[width=.5\columnwidth]{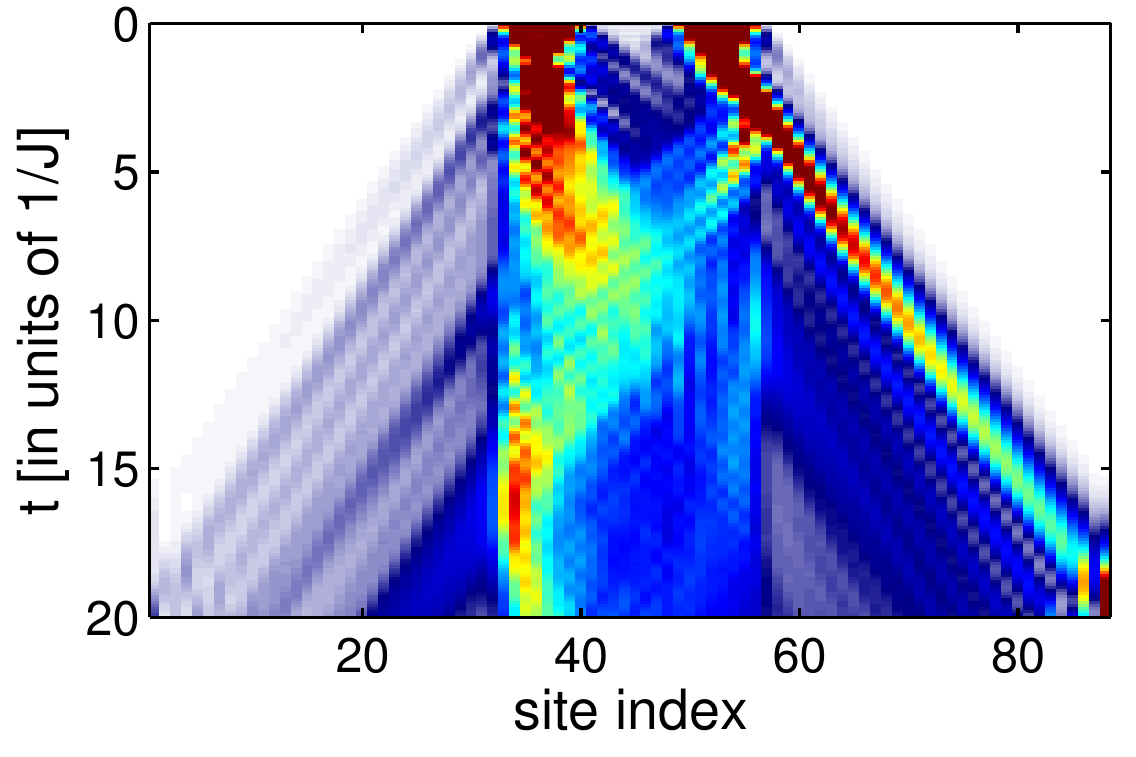}%
\includegraphics[width=.5\columnwidth]{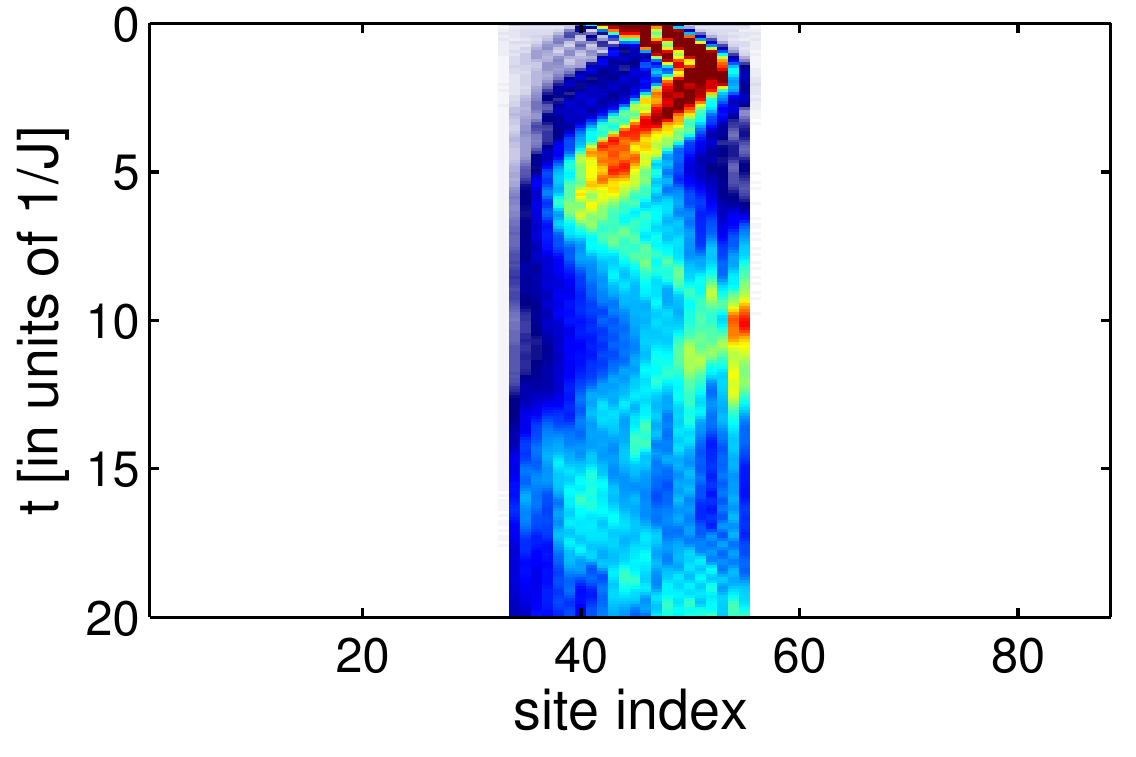}%
\caption{(Color online) 
Density of monomers (left column) and trimers (right column) 
in the $n=2$ MI cluster of 24 sites surrounded by empty lattice, 
$\tket{\cdot}{0}{32}$, on both sides. 
In the top panels, the initial state of the cluster
$\tket{-\pi/2}{2-}{8}\tket{\cdot}{2}{8} \tket{\pi/2}{2-}{8}$ 
corresponds to two monomers at the center of the band 
moving to the left and right.
In the central panel, the initial state
$\tket{\pi}{2-}{8}\tket{\cdot}{2}{8} \tket{0}{2-}{8}$
corresponds to two monomers at the upper and lower band edges. 
In the bottom panel, the initial state 
$\tket{\pi}{2-}{8}\tket{\pi/2}{2+}{8}\tket{0}{2-}{8}$
is the same as in the central panels plus a particle defect at the center 
of the band, moving to the right. The interaction strength is $U=100J$.
The density of monomers (trimers) corresponds to the probability of 
finding exactly one (three) particles at a given site. 
A TEBD \cite{Vidal2003} algorithm with bond dimension $\chi=200$ is used 
for the time evolution with a fourth order Trotter decomposition and 
time step size $1/50J$, with particle number conservation explicitly 
included in the MPS \cite{Daley2005a}.}
\label{fig:single_image}
\end{figure}

\begin{figure}[t]
\centering
\includegraphics[width=.5\columnwidth]{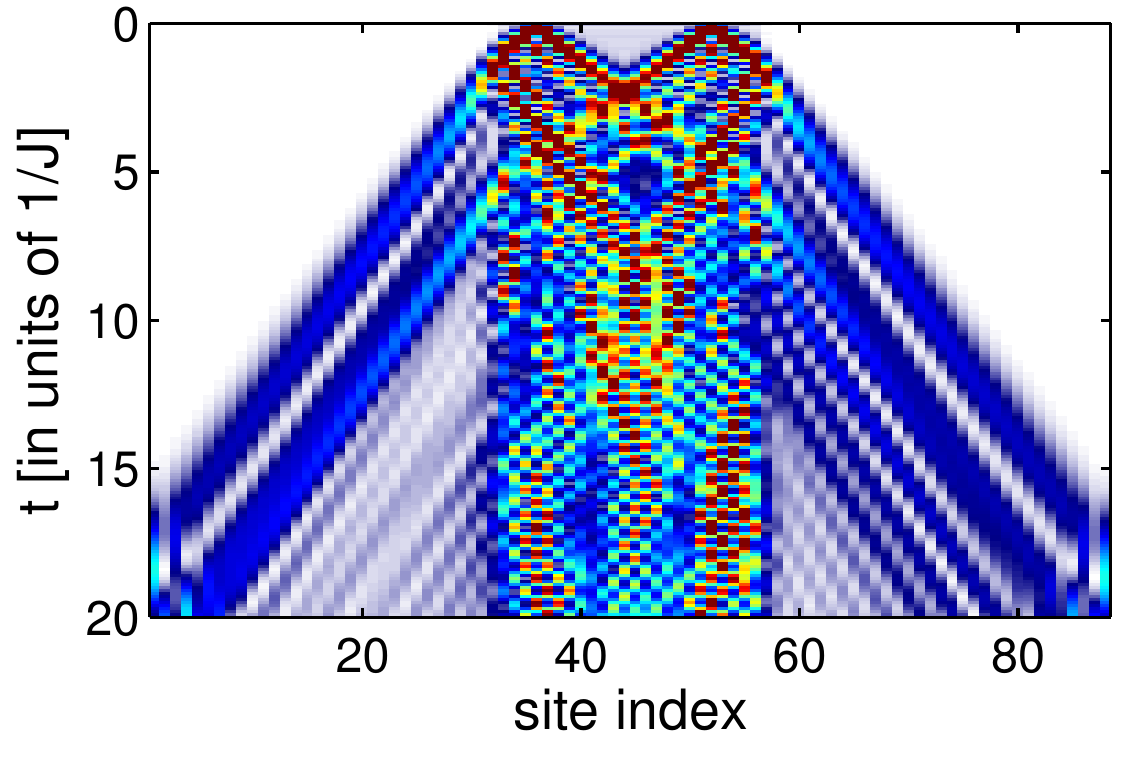}%
\includegraphics[width=.5\columnwidth]{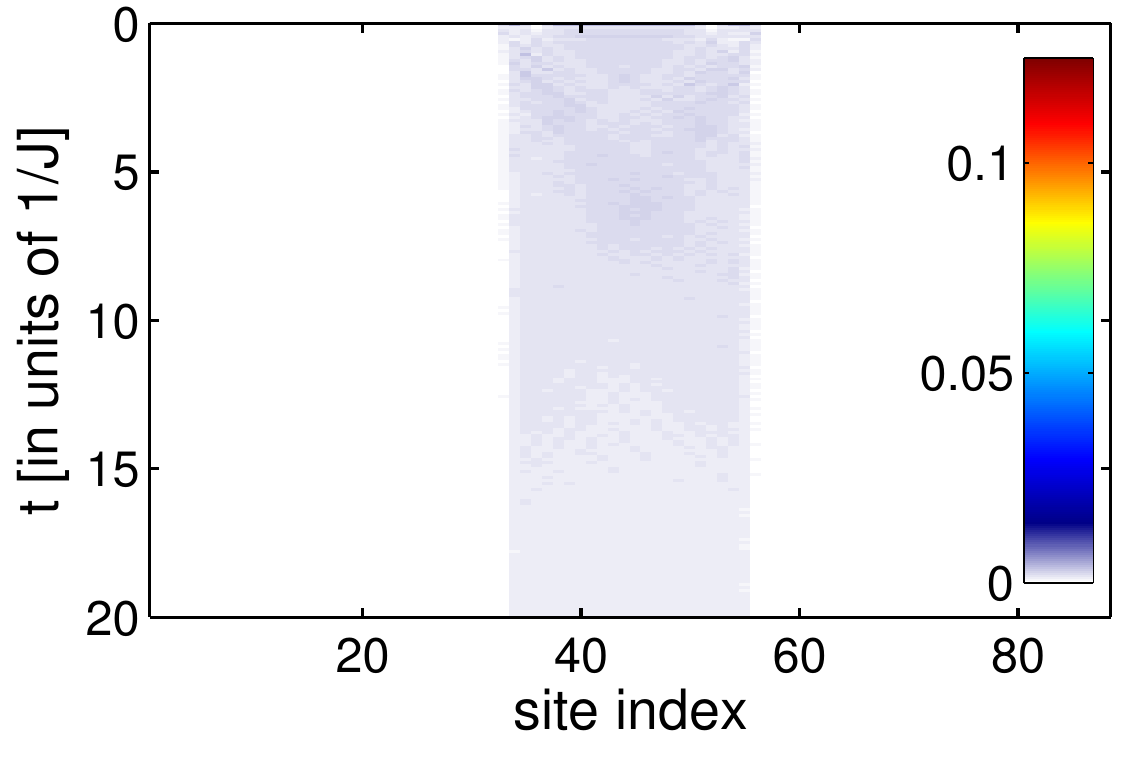}\\
\includegraphics[width=.5\columnwidth]{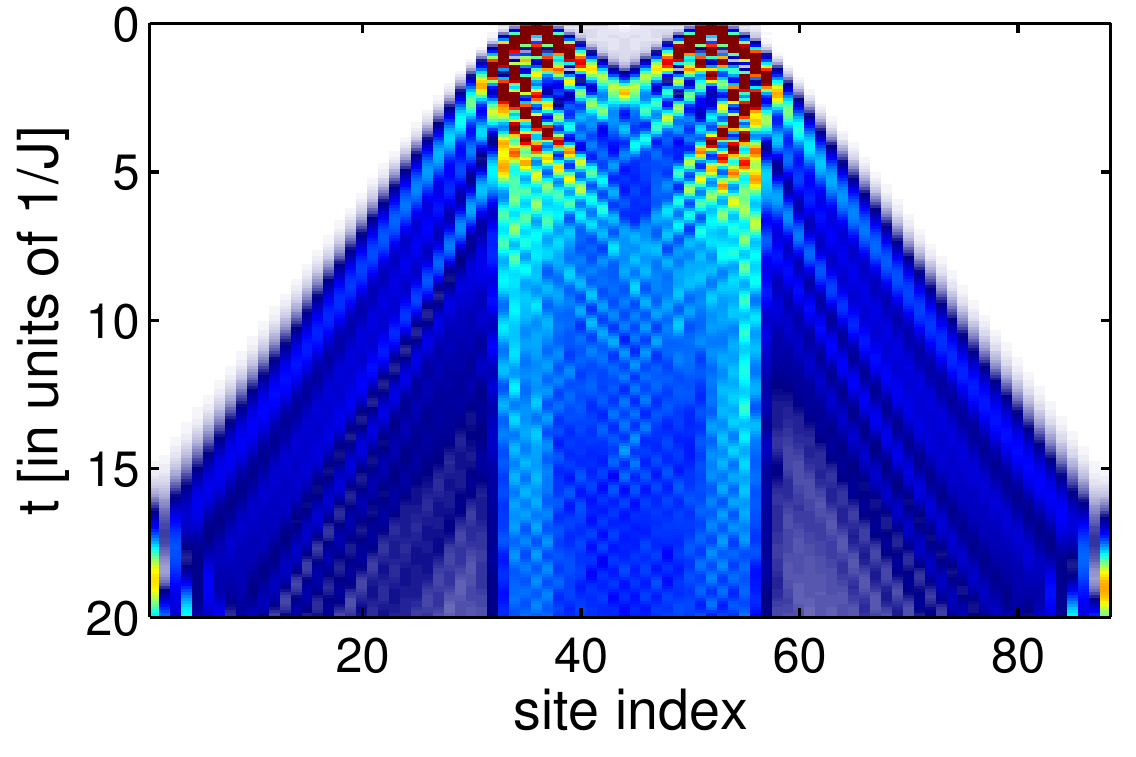}%
\includegraphics[width=.5\columnwidth]{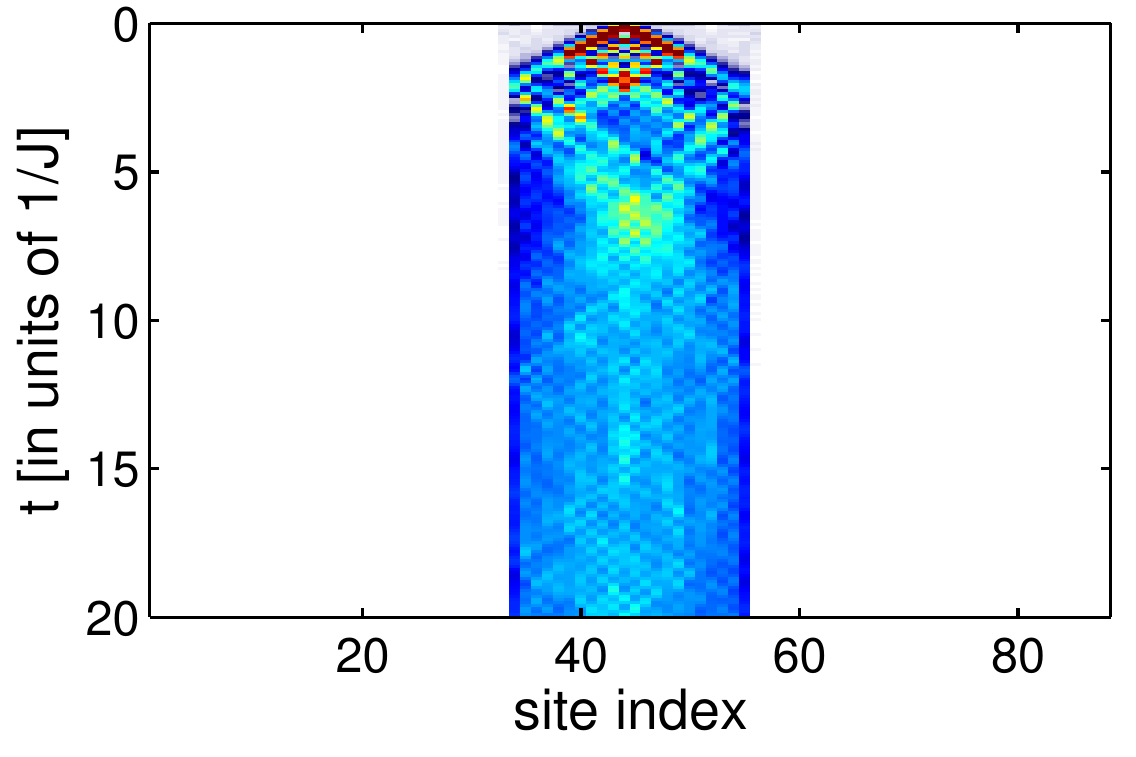}\\
\caption{(Color online) 
Same as Fig.~\ref{fig:single_image} but for localized initial states:
Top panel, $\tket{4_-}{2}{8} \tket{\cdot}{2}{8} \tket{4_-}{2}{8}$, 
two localized monomers;
bottom panel, $\tket{4_-}{2}{8} \tket{4_+}{2}{8} \tket{4_-}{2}{8}$, 
same plus a localized trimer.}
\label{fig:single_image_loc}
\end{figure}

Figures~\ref{fig:single_image} and \ref{fig:single_image_loc} show 
the time evolution of defects in a $n=2$ MI cluster surrounded by 
vacuum, obtained from the full BHM. 
Hole defects with quasi-momenta at the center of the band can easily leave 
the cluster after just a few scattering events, Fig.~\ref{fig:single_image}.
Hole defects prepared at the edges of the band remain trapped in the cluster.
An additional particle defect, which itself can not leave the cluster, 
induces fast quasi-momentum redistribution of the hole defects, large fraction 
of which can now leave the cluster.

The same effect is observed with localized defects, 
Fig.~\ref{fig:single_image_loc}. For hole defects alone, 
about one third of their population leaves the cluster (note that
the localized initial state of each defect has uniform distribution
of quasi-momentum $k \in [-\pi, \pi]$, and not energy $E_k \in [-4J,4J]$), 
while an additional localized particle defect increases this fraction 
significantly. 

\begin{figure}[b]
\centering
a)\includegraphics[width=.6\columnwidth]{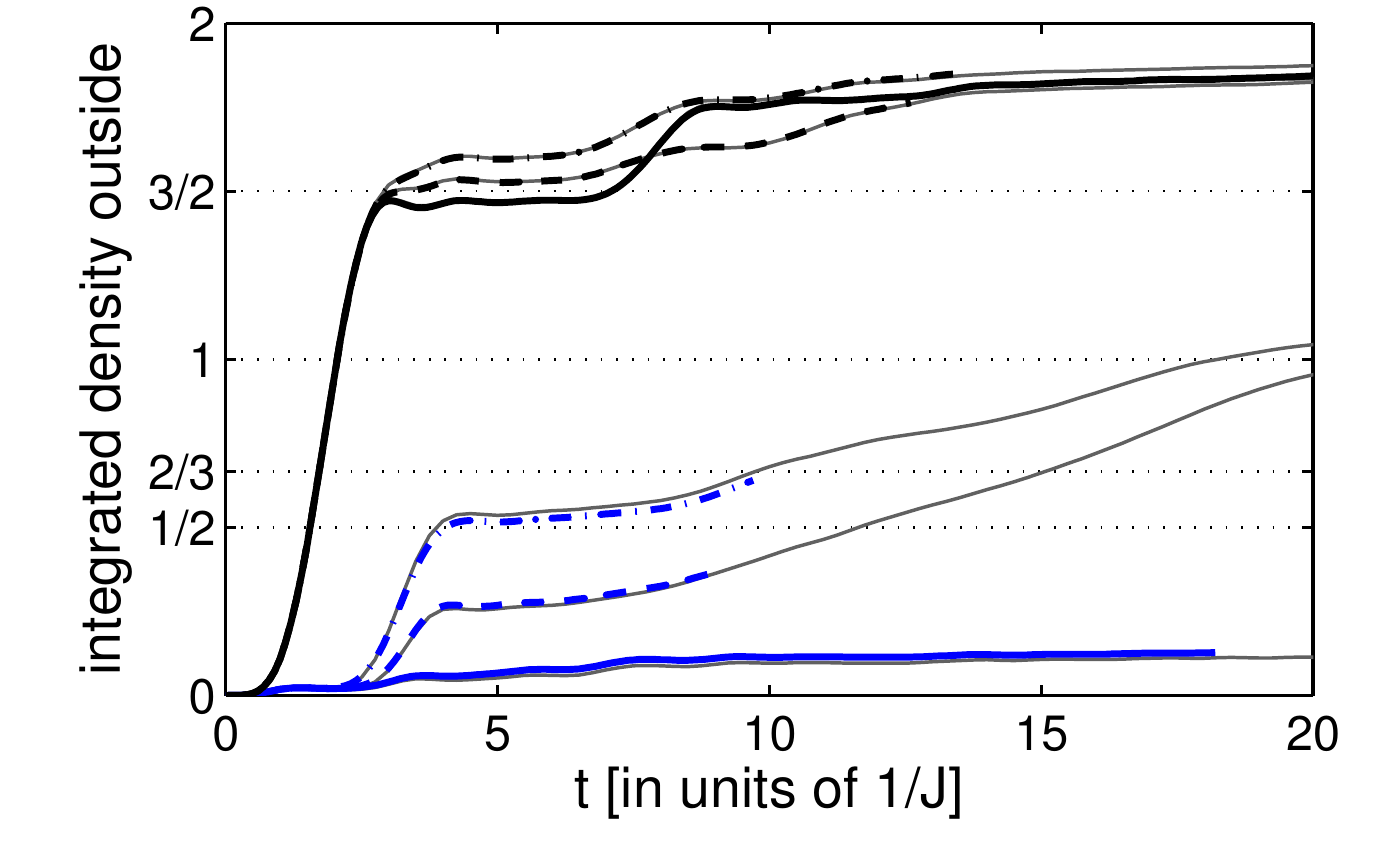}\\
b)\includegraphics[width=.6\columnwidth]{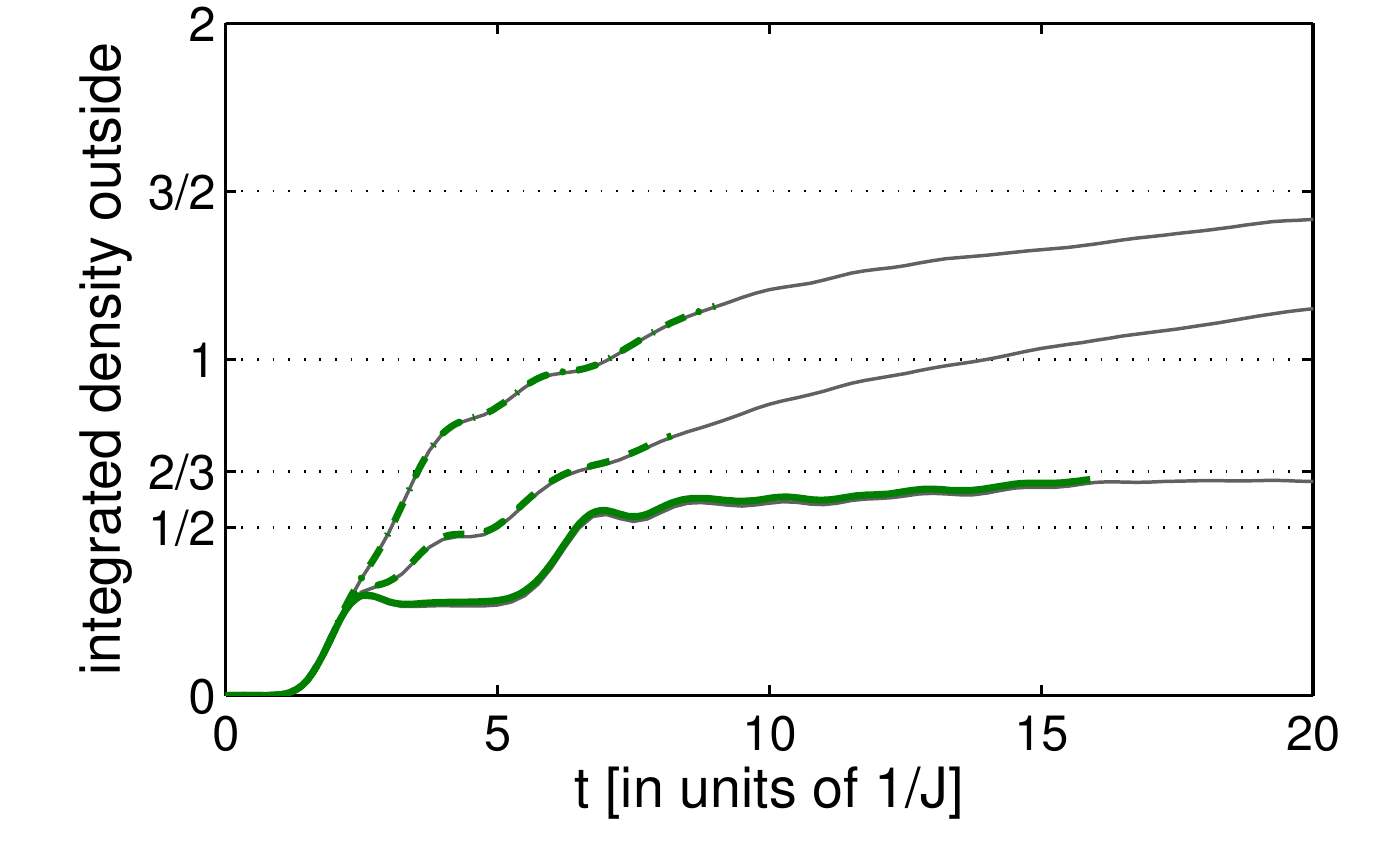}
\caption{(Color online) Total population 
(integrated density,  $\sum_{j=1}^{31} \expv{\had_j\ha_j} 
+ \sum_{j=58}^{88} \expv{\had_j\ha_j}$) of monomers outside the dimer cluster. 
(a) Initial states of the cluster are: in the upper black branch,
$\tket{-\pi/2}{2-}{8}\tket{\cdot}{2}{8} \tket{\pi/2}{2-}{8}$ (solid line),
$\tket{-\pi/2}{2-}{8}\tket{4_+}{2}{8} \tket{\pi/2}{2-}{8}$ (dashed line), and
$\tket{-\pi/2}{2-}{8}\tket{\pi/2}{2+}{8} \tket{\pi/2}{2-}{8}$ (dot-dashed line);
in the lower blue branch,
$\tket{\pi}{2-}{8}\tket{\cdot}{2}{8} \tket{0}{2-}{8}$ (solid line),
$\tket{\pi}{2-}{8}\tket{4_+}{2}{8} \tket{0}{2-}{8}$ (dashed line), and
$\tket{\pi}{2-}{8}\tket{\pi/2}{2+}{8} \tket{0}{2-}{8}$ (dot-dashed line).
(b) Initial states of the cluster are: 
$\tket{4_-}{2}{8} \tket{\cdot}{2}{8} \tket{4_-}{2}{8}$ (solid line),
$\tket{4_-}{2}{8} \tket{4_+}{2}{8}  \tket{4_-}{2}{8}$ (dashed line), and
$\tket{4_-}{2}{8} \tket{\pi/2}{2+}{8} \tket{4_-}{2}{8}$ (dot-dashed line).
 The interaction strength is $U=100J$. Numerical parameters are the same 
as in Figs.~\ref{fig:single_image} and \ref{fig:single_image_loc}, 
and the curves terminate when the accumulated 
cut-off error equals $10^{-2}$. The gray lines are obtained from the
equivalent effective model, with the time step increased to $1/10J$.}
\label{fig:single_integrated_density}
\end{figure}

In Fig.~\ref{fig:single_integrated_density}(a) and (b) we show 
the time evolution of the total population of monomers outside 
the dimer cluster pertaining to the cases illustrated in  
Figs.~\ref{fig:single_image} and \ref{fig:single_image_loc}, respectively.
Again, hole defects with quasi-momenta in the center of the band easily
escape the cluster even without the assistance of a particle defect,
Fig.~\ref{fig:single_integrated_density}(a). Conversely, for the hole 
defects with quasi-momenta at the edges of the Bloch band in the cluster, 
very little population is found outside the cluster in the long time limit
(the small fraction of monomer population in the vacuum is due to the finite
binding energy $U$ of the dimers). Adding a particle defect in the cluster 
significantly increases the fraction of monomers outside the cluster; 
we find that the increase is always larger for a particle defect in 
the center of the band than for a localized one.

For the initially localized hole defects,
Fig.~\ref{fig:single_integrated_density}(b), and without assistance of
a particle defect, we find that, as expected, about a third of their 
total population occupying the center of the Bloch band leaves the cluster 
in the long-time limit. A particle defect can further increase the portion
of escaping population of the hole defects by redistributing their 
quasi-momenta over the entire band.  

Note that the results of numerical simulations for the system with a particle 
defect are reliable for shorter times as compared to the simulations with 
the hole defects only, which is due to the larger entanglement created 
dynamically upon the trimer-monomer collisions. 

So far we have been restricted to the treatment of only two monomers 
and one trimer and for relatively short times, because in the full BHM
the fast growing entanglement in the system limits the numerical method.
With the effective model containing only the hole and particles defects,
we can simulate the dynamics for much longer times with the same numerical 
accuracy, as can be seen in Fig. \ref{fig:single_integrated_density}. 
The perfect agreement between the full and effective models 
allows us to employ the effective model for simulating larger 
systems and for longer times. 

\begin{figure}[t]
\centering
a)\includegraphics[width=.6\columnwidth]{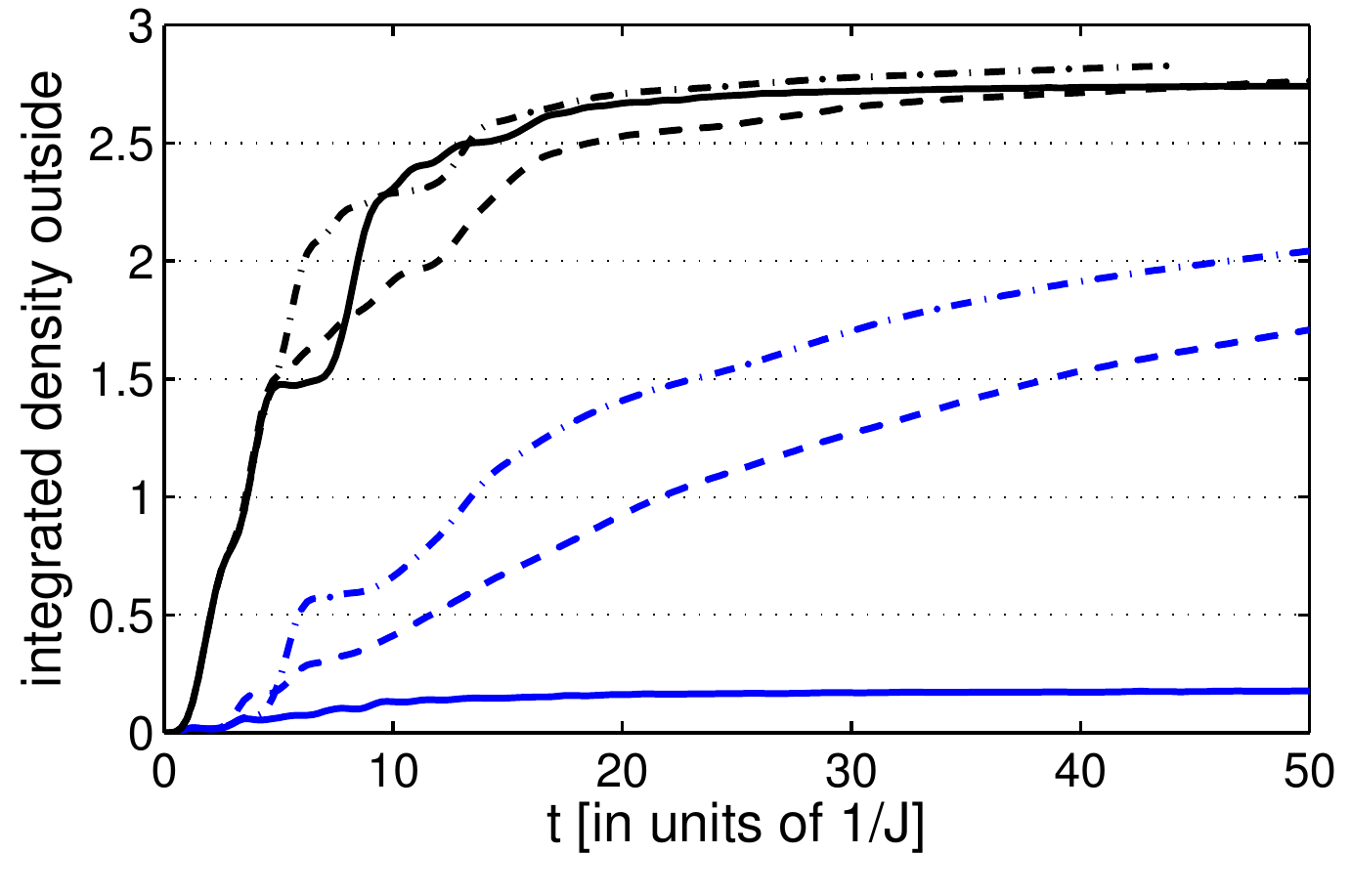}\\
b)\includegraphics[width=.6\columnwidth]{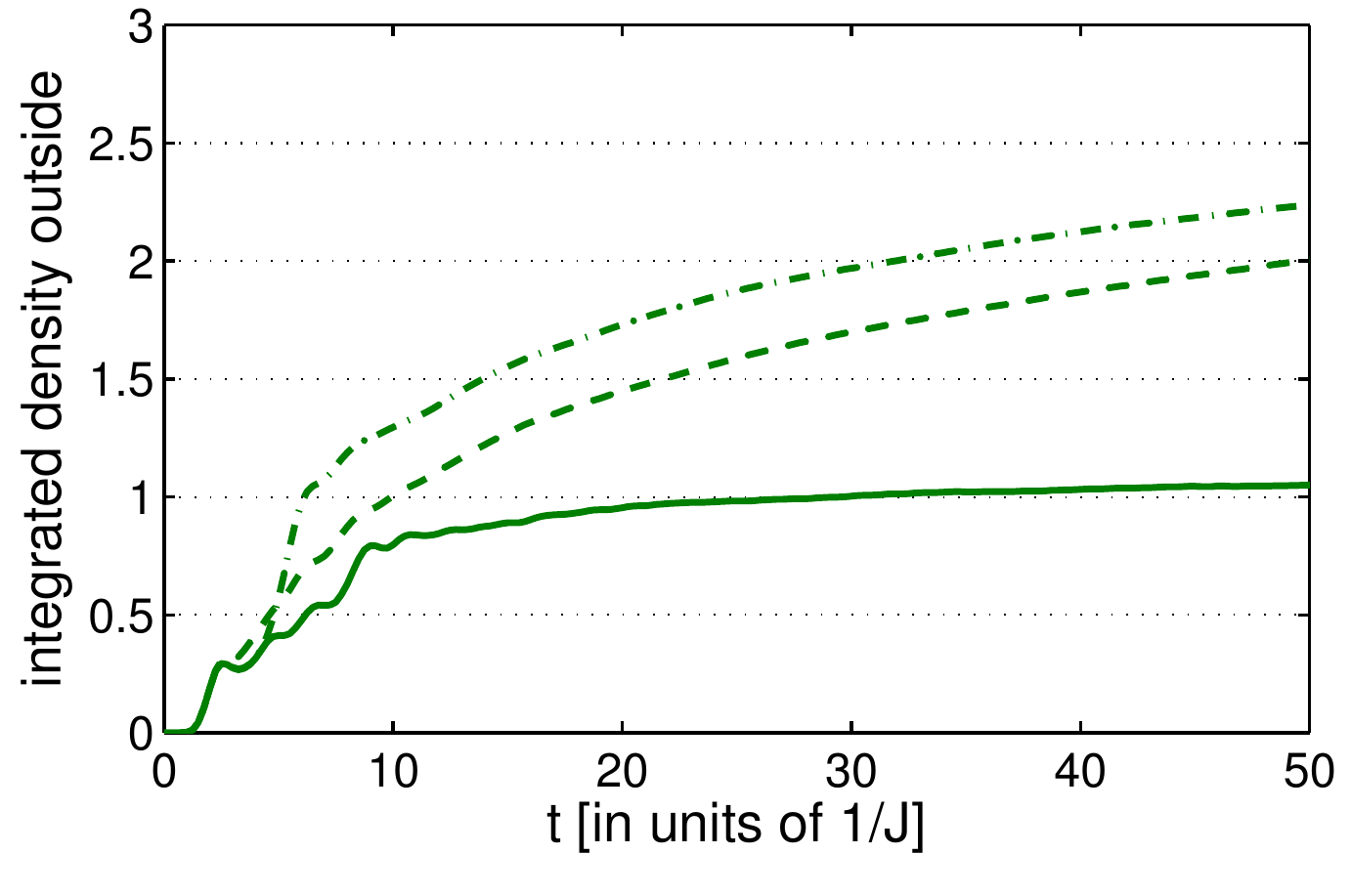}\\
c)\includegraphics[width=.6\columnwidth]{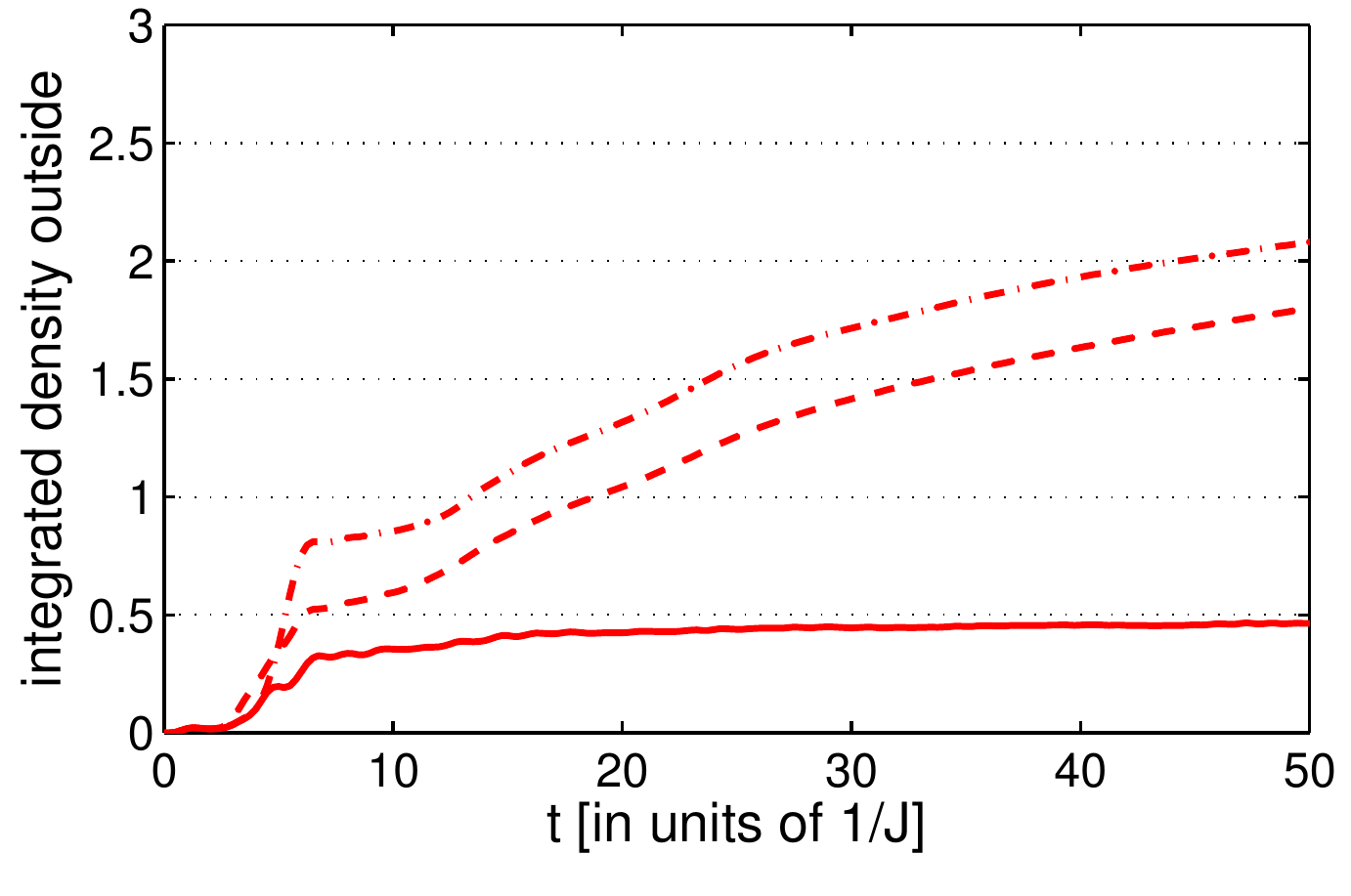}
\caption{(Color online) Total population 
($\sum_{j=1}^{63} \expv{\had_j\ha_j} + \sum_{j=98}^{160}\expv{\had_j\ha_j}$) 
of monomers outside the dimer cluster of 32 sites surrounded by empty lattice, 
$\tket{\cdot}{0}{64}$, on both sides.
(a) Initial states of the cluster are: in the upper black branch,
$\tket{-\pi/2}{2-}{8} \tket{\cdot}{2}{8} 
 \tket{\pi/2}{2-}{8} \tket{-\pi/2}{2-}{8}$ (solid line),
$\tket{-\pi/2}{2-}{8}\tket{4_+}{2}{8} 
 \tket{\pi/2}{2-}{8} \tket{-\pi/2}{2-}{8}$ (dashed line), and
$\tket{-\pi/2}{2-}{8}\tket{\pi/2}{2+}{8} 
 \tket{\pi/2}{2-}{8} \tket{-\pi/2}{2-}{8}$ (dot-dashed line);
in the lower blue branch,
$\tket{\pi}{2-}{8}\tket{\cdot}{2}{8}
 \tket{0}{2-}{8} \tket{\pi}{2-}{8}$ (solid line),
$\tket{\pi}{2-}{8}\tket{4_+}{2}{8}
 \tket{0}{2-}{8}\tket{\pi}{2-}{8}$ (dashed line), and
$\tket{\pi}{2-}{8}\tket{\pi/2}{2+}{8}
 \tket{0}{2-}{8}\tket{\pi}{2-}{8}$ (dot-dashed line).
(b) Initial states of the cluster are: 
$\tket{4_-}{2}{8}\tket{\cdot}{2}{8}
 \tket{4_-}{2}{8}\tket{4_-}{2}{8}$ (solid line),
$\tket{4_-}{2}{8}\tket{4_+}{2}{8}
 \tket{4_-}{2}{8}\tket{4_-}{2}{8}$ (dashed line), and
$\tket{4_-}{2}{8} \tket{\pi/2}{2+}{8}
 \tket{4_-}{2}{8}\tket{4_-}{2}{8}$ (dot-dashed line).
(c) Initial states of the cluster are: 
$\tket{\pi}{2-}{8}\tket{\cdot}{2}{8}
 \tket{0}{2-}{8}\tket{4_-}{2}{8}$ (solid line),
$\tket{\pi}{2-}{8}\tket{4_+}{2}{8}
 \tket{0}{2-}{8}\tket{4_-}{2}{8}$ (dashed line), and
$\tket{\pi}{2-}{8} \tket{\pi/2}{2+}{8}
 \tket{4_-}{2}{8}\tket{4_-}{2}{8}$ (dot-dashed line).
Simulations were performed with the effective model.
Bond dimensions $\chi=300$ are used, and the time step size is $1/10J$. 
The curves terminate when the accumulated cut-off error equals $10^{-1}$.}
\label{fig:single_integrated_density643264}
\end{figure}

Fig. \ref{fig:single_integrated_density643264} shows numerical results for 
a system containing initially up to four defects. As expected, the evaporation
works for the larger systems as well. Most importantly, in the presence of
a particle defects, the number of hole defects left in the cluster
in the long-time limit falls well below unity (extrapolating 
the curves to somewhat larger times than shown in 
Fig.~\ref{fig:single_integrated_density643264}, if necessary).

\section{Two species Bose-Hubbard model}
\label{sec:twospecies}

We have seen in the previous sections that, in the single species 
BHM, the hopping amplitudes of a monomer inside an $n=2$ MI cluster 
and on an empty lattice differ by a fixed factor of 2. 
More flexibility is offered by the two species BHM, which we now 
briefly discuss. The Hamiltonian for the system is
\begin{eqnarray}
 \hH &=& - \Ja \sum_j (\had_j\ha_{j+1} + \hadj) 
- \Jb \sum_j (\hbd_j\hb_{j+1} + \hadj )
\nonumber \\ & & 
+ \frac{\Ua}2 \sum_j \had_j\had_j\ha_j\ha_j
+ \frac{\Ub}2 \sum_j \hbd_j\hbd_j\hb_j\hb_j 
\nonumber \\ & &  
+ \Uab\sum \had_j\ha_j\hbd_j\hb_j, \label{eq:twospeciesham}
\end{eqnarray}
where $\ha_j$ ($\hb_j$) are the bosonic operators for the particles
of type $\mathrm{a}$ ($\mathrm{b}$) hopping between adjacent sites with 
the rate $\Ja$ ($\Jb$), while $\Ua,\Ub$ and $\Uab$ are the intra- and 
inter-species on-site interactions.  

Assuming the conditions $\Ua, \Ub, \Uab, |\Ua+\Ub-2\Uab| \gg \Ja, \Jb$,
we first  consider the situation where each lattice site is either empty 
or contains a single $\mathrm{a}$-$\mathrm{b}$ dimer, that is, a pair of 
strongly interacting (via $\Uab$) particles $\mathrm{a}$ and $\mathrm{b}$ 
localized on the same site. Upon adiabatic elimination of the non resonant
states containing unpaired particles on neighboring sites \cite{Schmidt2009a},
we obtain an effective Hamiltonian of the form of Eq.~(\ref{eq:Hpair}),
where now the dimer hopping and nearest-neighbor interaction are given by
\begin{equation}
\tJ = - \frac{2 \Ja \Jb}{\Uab} , \;\;
\tB = -2 \left( \frac{2\Ja^2}{\Ua} + \frac{2\Jb^2}{\Ub} + 
\frac{\Ja^2 + \Jb^2}{\Uab} \right) .
\end{equation}
With all the interactions repulsive, the anisotropy parameter
\begin{equation}
 \Delta = \tB/2\tJ = \frac{\Ja}{\Jb} \left(\frac12 + \frac{\Uab}{\Ua}\right) 
+ \frac{\Jb}{\Ja} \left(\frac12+\frac{\Uab}{\Ub}\right) 
\end{equation}
is larger than $1$ for any finite $\Ua/\Uab$ or $\Ub/\Uab$, 
and the MI cluster of $\mathrm{a}$-$\mathrm{b}$ dimers is stable. 
But for $\Ua/\Uab, \Ub/\Uab \rightarrow \infty$, corresponding to 
the band insulator for two fermionic species, $\Delta = 1$ and 
the dimer cluster is unstable. 
 

Inside the $n = \na + \nb= 2$ ($\na=\nb=1$) MI cluster, a hole defect 
of type $\mathrm{a}$ (unpaired particle $\mathrm{b}$) is created by $\ha_j$, 
see Fig. \ref{fig:particles}(e). The defect hops in the cluster 
with the rate $\Ja$ while outside the cluster its hopping rate is $\Jb$.
It must be stable and not resonantly converted into a pair of 
particles $\mathrm{b}$ and a single $\mathrm{b}$-hole (unpaired 
particle $\mathrm{a}$), which requires that $\Ub -\Uab \gg \Ja, \Jb$. 
Neglecting the second-order corrections of the order of 
$J_{\rm a,b}^2/U_{\rm a,b,ab}$, we have the effective single-particle 
Hamiltonian (\ref{eq:HTrans}) with $\JA = \Ja$ and $\JB = \Jb$.
Using the results of Appendix~\ref{sec:apptrans}, we calculate the 
transmission probability $T(k)$ of the particle through the domain 
wall separating the regions A and B for various $\Jb/\Ja$, which 
is shown in Fig. \ref{fig:Talphak}. At $\Ja=\Jb$ ($\alpha =1$) 
we find an almost perfect transmission for all $k$, up to 
a small correction due to finite interactions.
The above results equally apply to a hole defect of type $\mathrm{b}$ 
(unpaired particle $\mathrm{a}$) with the replacement 
$\mathrm{a} \leftrightarrow \mathrm{b}$.


We have performed numerical simulations of the dynamics of several defects 
in a dimer cluster surrounded by vacuum using the full model of 
Eq.~(\ref{eq:twospeciesham}). For computational reasons, we truncate the 
local Hilbert space to three bosons of each species per site, which is justified
by the facts that, due to the strong interactions, the occupation of 
a single site by more particles can safely be neglected. 

\begin{figure}[t]
\centering
\includegraphics[width=.5\columnwidth]{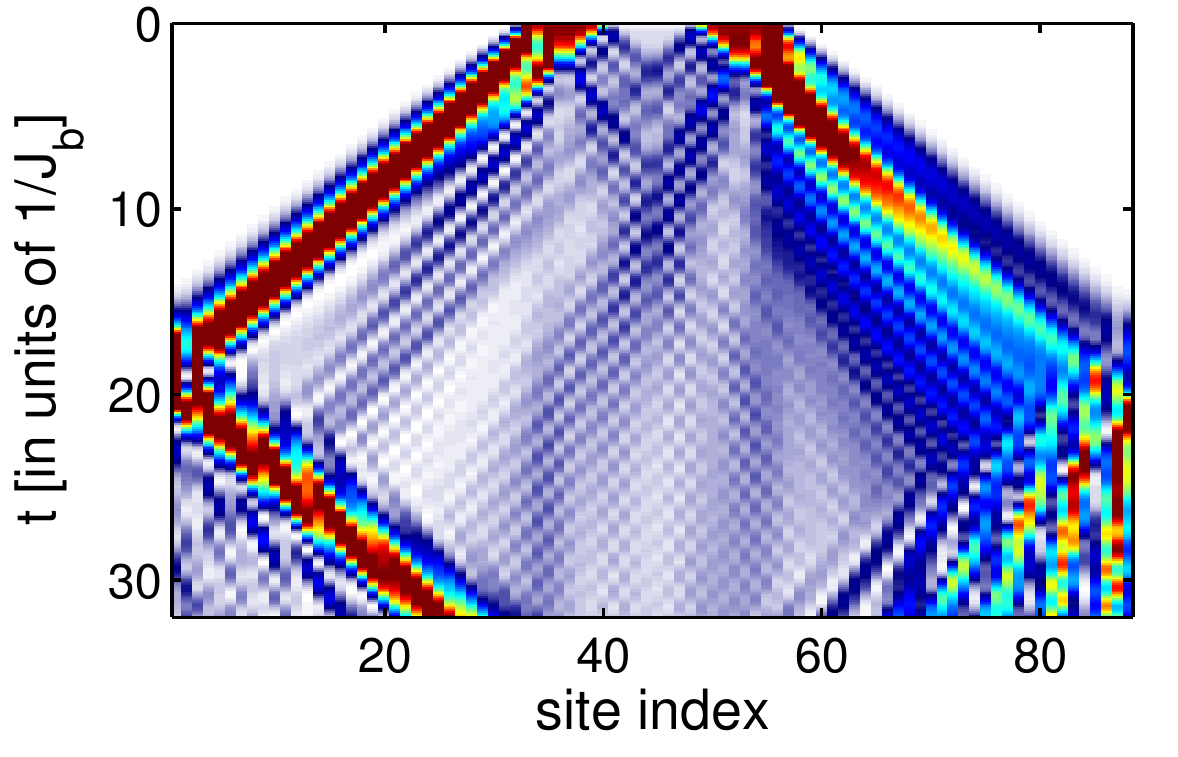}%
\includegraphics[width=.5\columnwidth]{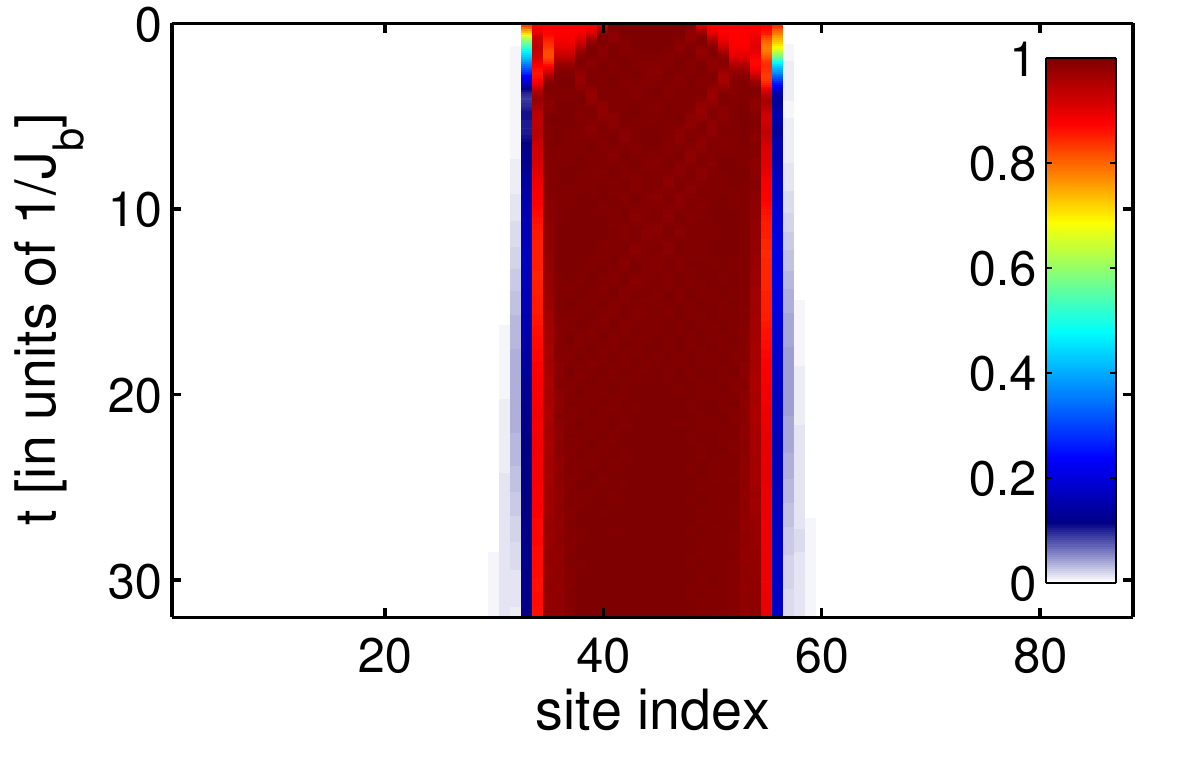}\\
\includegraphics[width=.5\columnwidth]{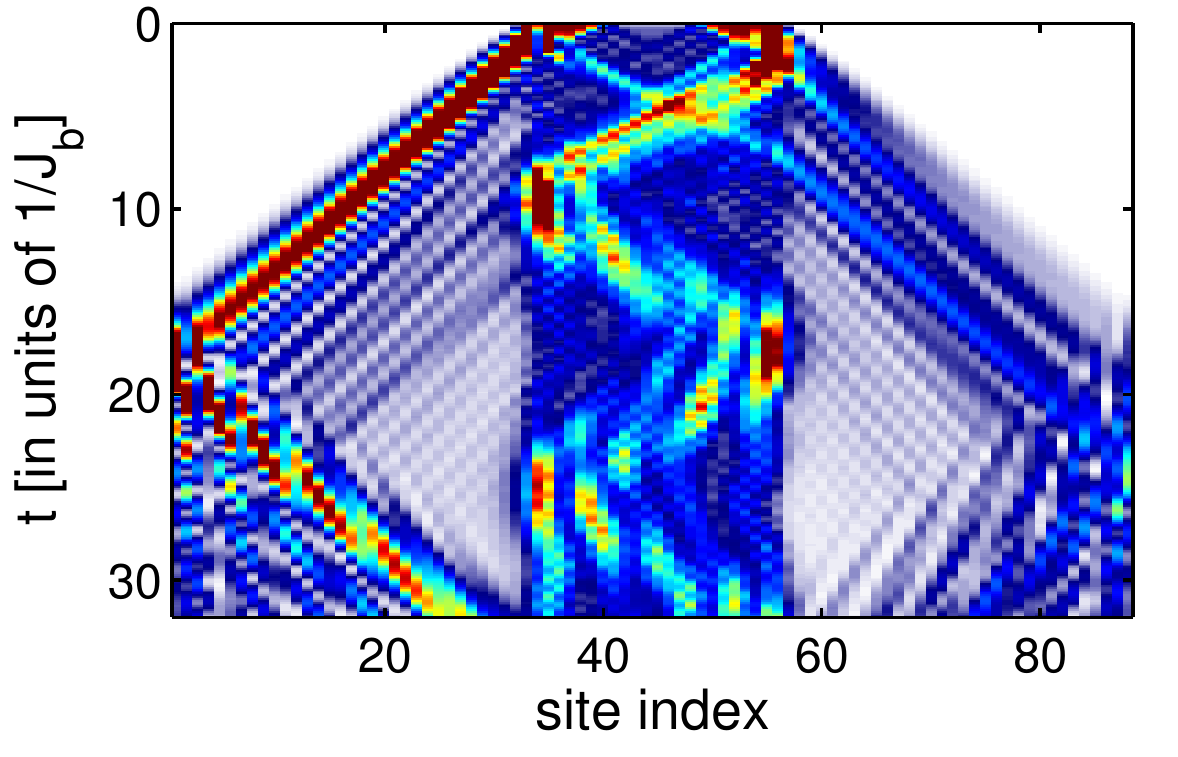}%
\includegraphics[width=.5\columnwidth]{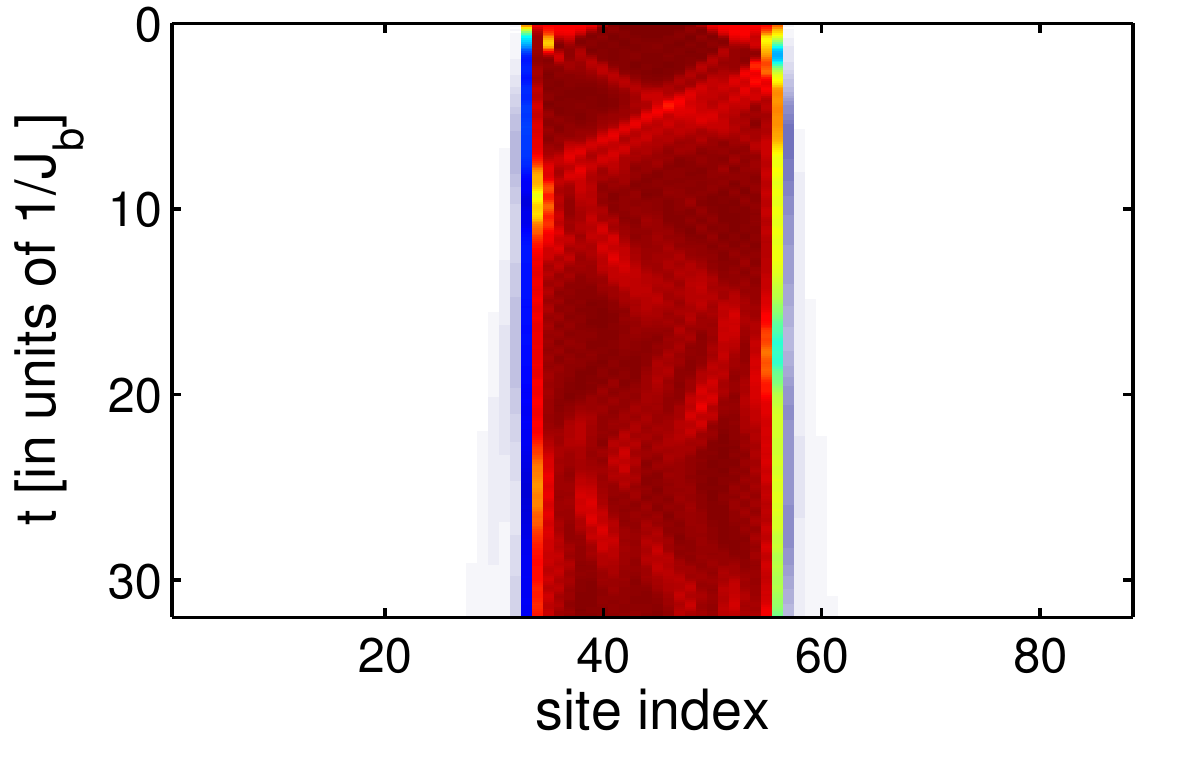}\\
\includegraphics[width=.5\columnwidth]{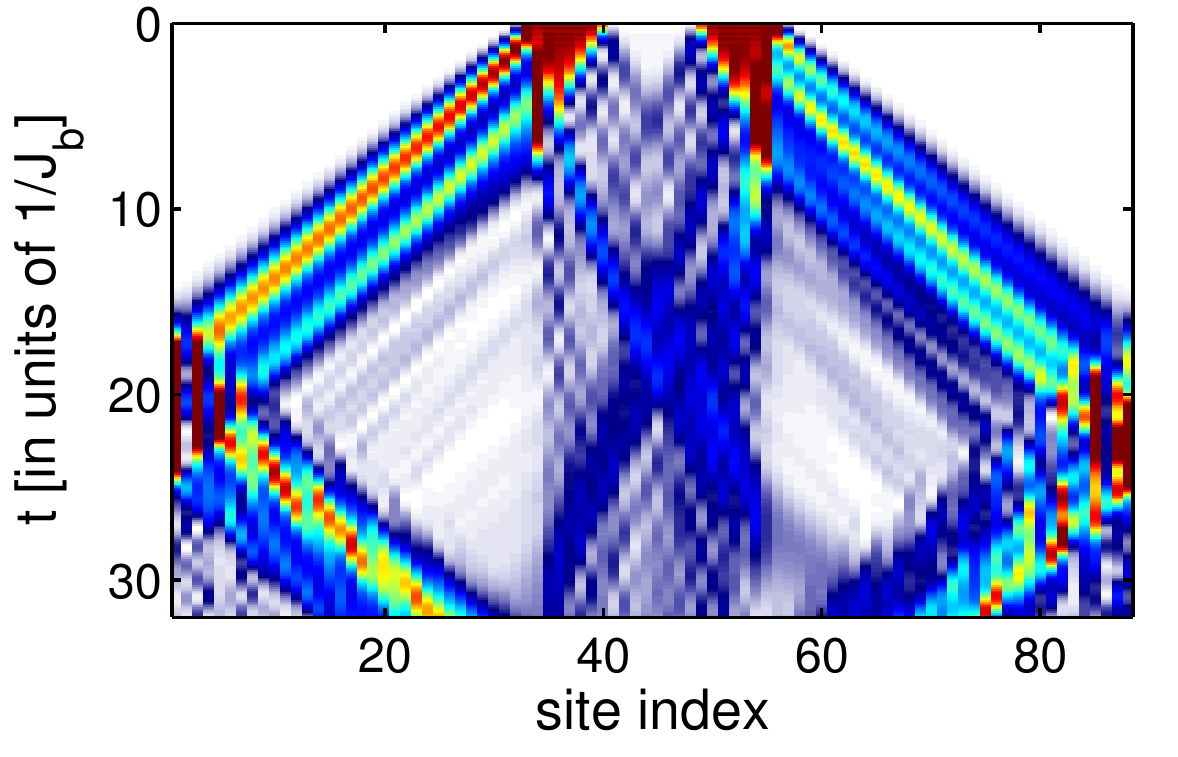}%
\includegraphics[width=.5\columnwidth]{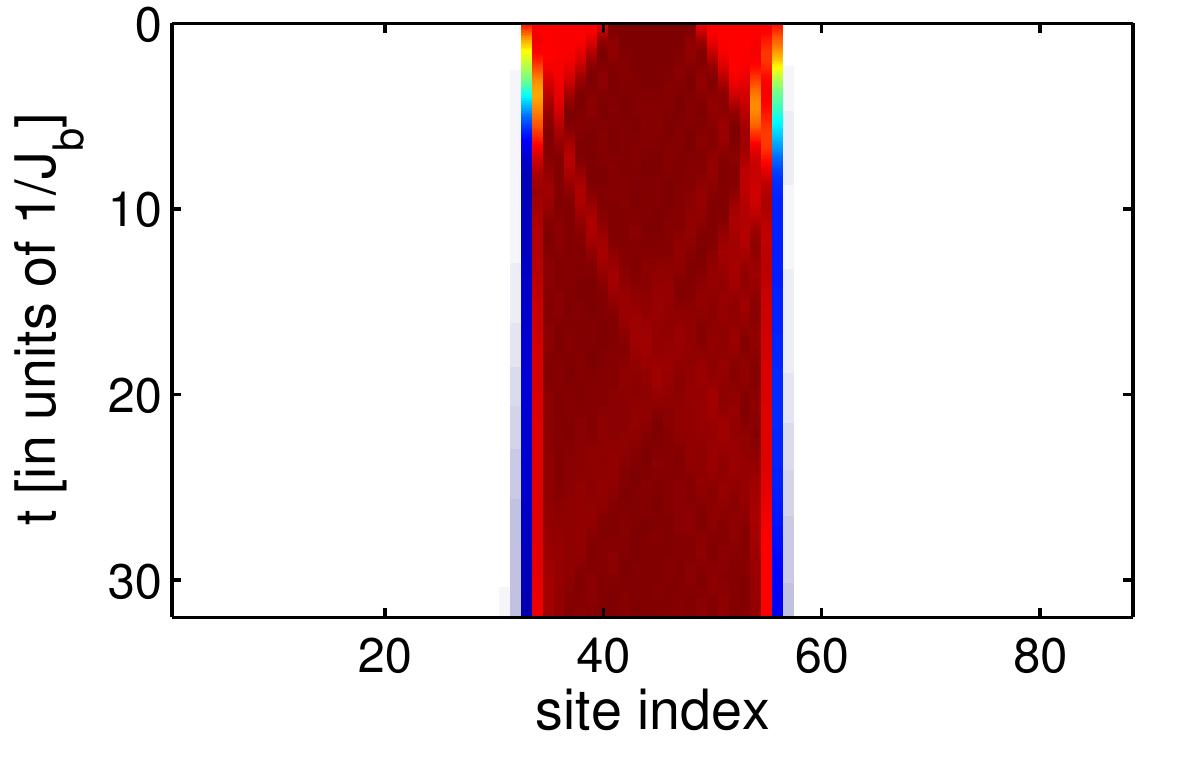}%
\caption{(Color online) 
Density of unpaired particles $\mathrm{b}$, or $\mathrm{a}$-holes,
in the cluster (left column), and particles $\mathrm{a}$ (right column),
in the lattice with a MI cluster of $\mathrm{a}$-$\mathrm{b}$ dimers 
spanning 24 sites surrounded by empty lattice, $\tket{\cdot}{0}{32}$, 
on both sides. The initial state of the cluster 
$\tket{-\pi/2}{2-\mathrm{a}}{8}\tket{\cdot}{2}{8}\tket{\pi/4}{2-\mathrm{a}}{8}$
corresponds to two $\mathrm{a}$-hole defects moving to the left with velocity 
$2\Ja$, and to the right with velocity $\sqrt{2} \Ja$, respectively, 
while all particles $\mathrm{a}$ are dimerized with particles $\mathrm{b}$ 
(no $\mathrm{b}$-hole defects). 
The parameters are $\Ua=\Ub=60\Jb$, $\Uab=40\Jb$, and $\Ja = \Jb$ (top panel),
$\Ja = 2\Jb$ (central panel), and $\Ja = \frac12\Jb$ (bottom panel). 
A TEBD algorithm with bond dimension $\chi=100$ is used for the time 
evolution with a fourth order Trotter decomposition and time step size 
$1/50\Jb$, with the particle number conservation for each species
explicitly included in the MPS.}
\label{fig:two_image_mfq_fff}
\end{figure}

In Fig.~\ref{fig:two_image_mfq_fff} we show the behavior of two 
unpaired particles $\mathrm{b}$, or $\mathrm{a}$-holes, moving 
in the cluster with different initial velocities. 
In the case of $\Ja = \Jb$ (top panel), both defects almost completely 
leave the cluster upon the first encounter with its walls. 
For $\Ja \neq \Jb$, only partial transmission of each defects 
is recorded, which depends on its initial quasi-momentum, as per
Fig. \ref{fig:Talphak}. As an example, at $\Ja = 2\Jb$ 
(central panel) the unpaired particle b with $k=\pi/2$ can leave 
the cluster, while that with $k=\pi/4$ can not, as its quasi-momentum
is close to the lower band edge.

\begin{figure}[t]
\centering
\includegraphics[width=.5\columnwidth]{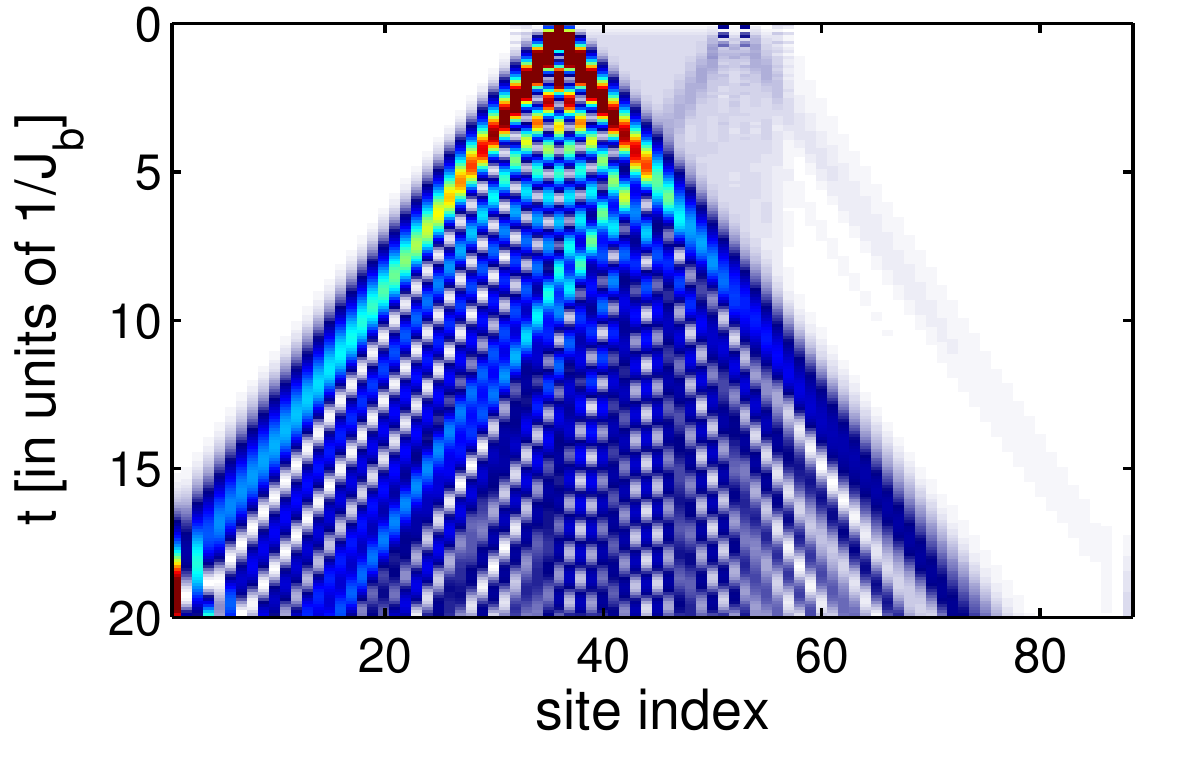}%
\includegraphics[width=.5\columnwidth]{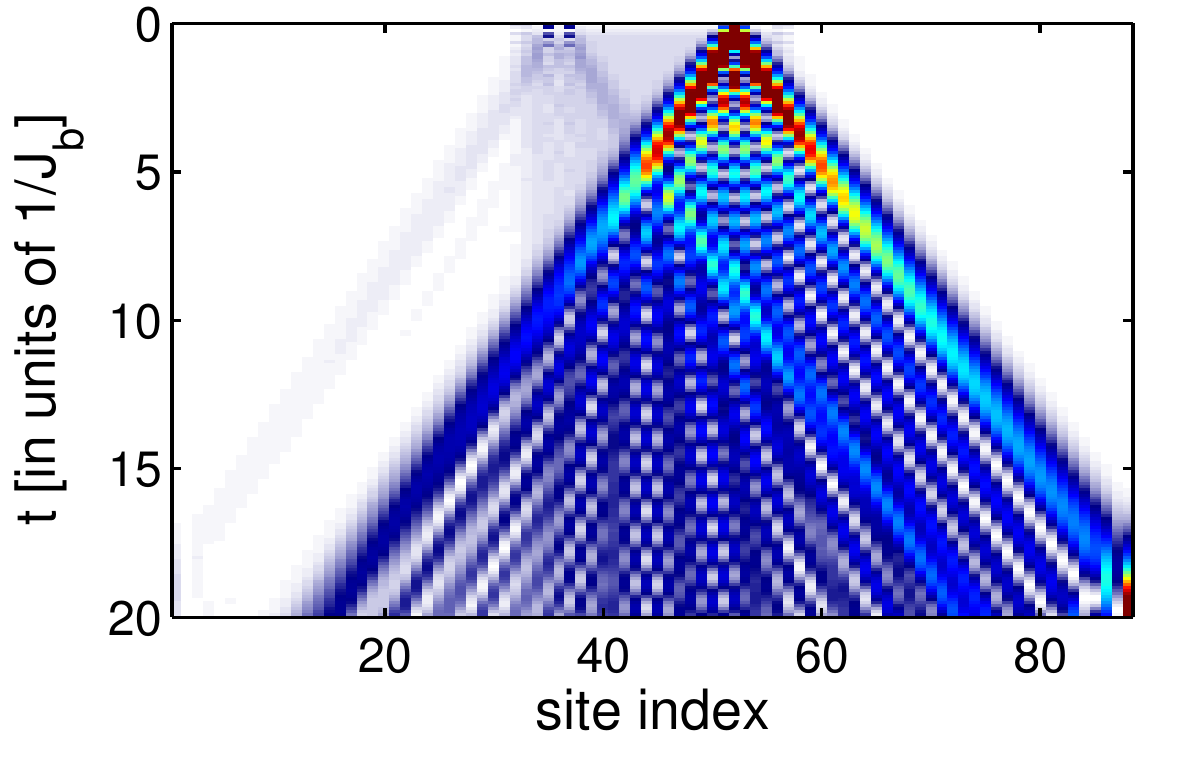}\\
\includegraphics[width=.5\columnwidth]{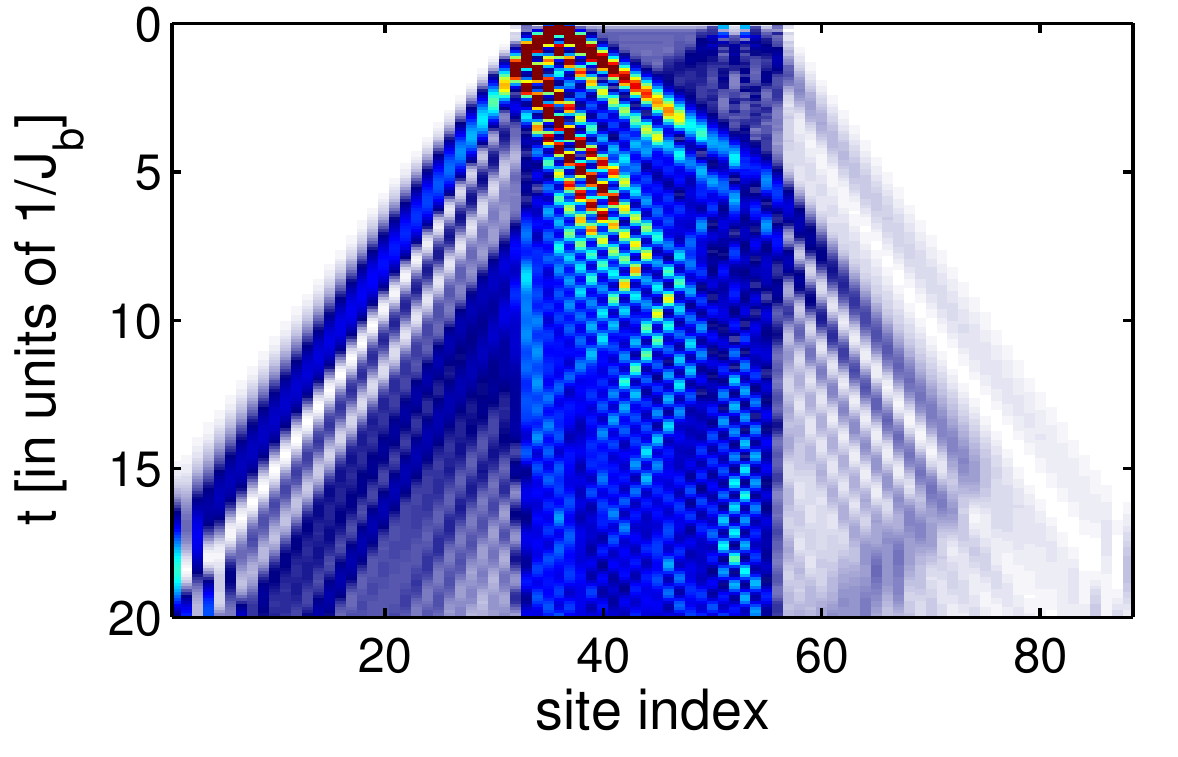}%
\includegraphics[width=.5\columnwidth]{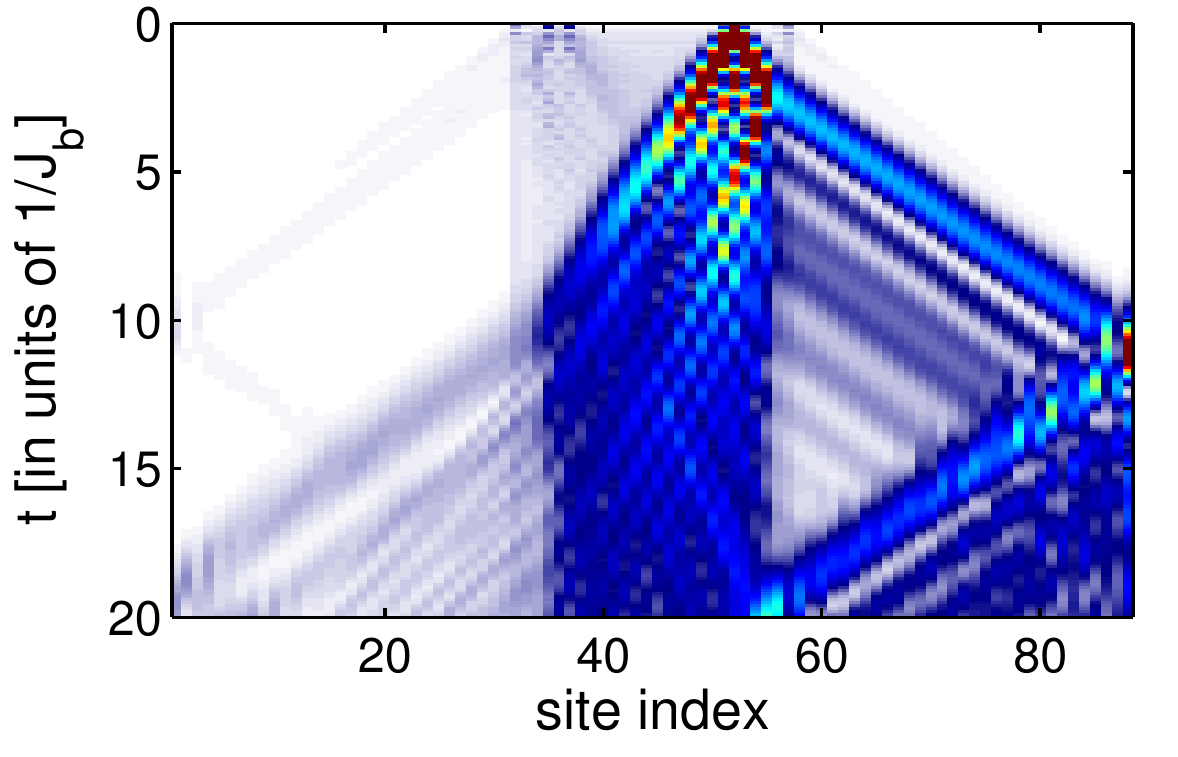}\\
\includegraphics[width=.5\columnwidth]{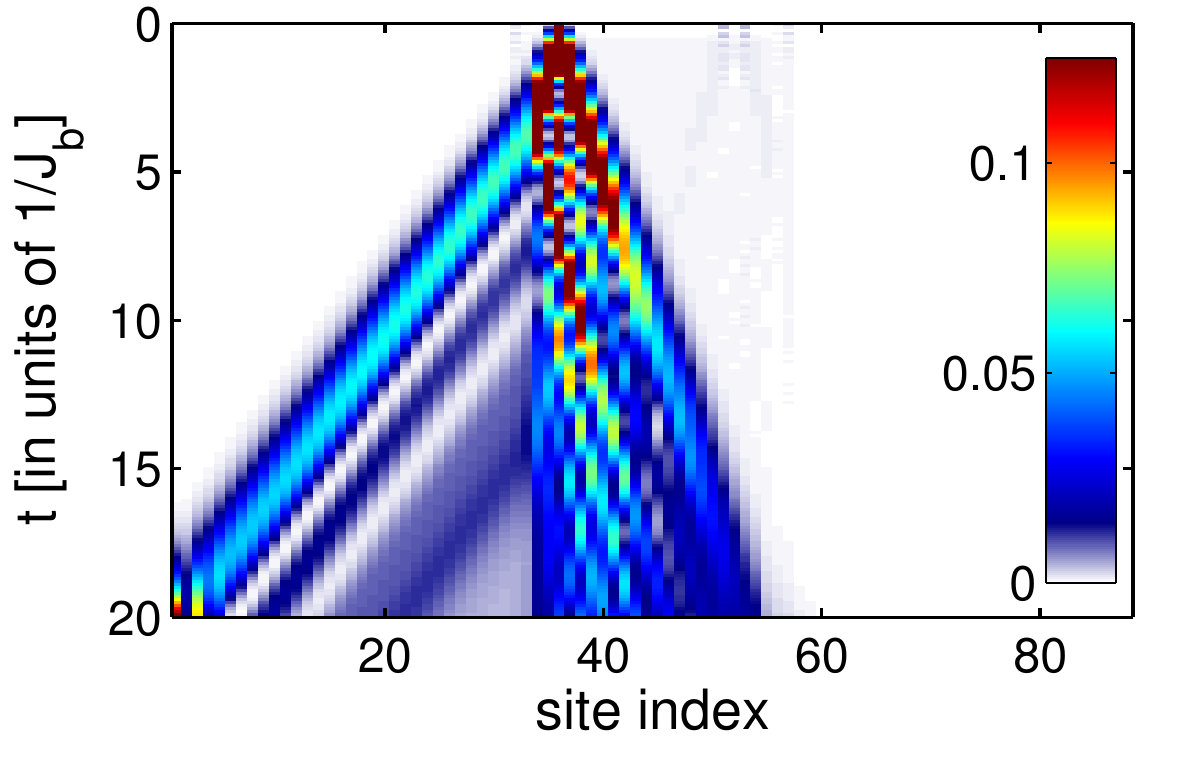}%
\includegraphics[width=.5\columnwidth]{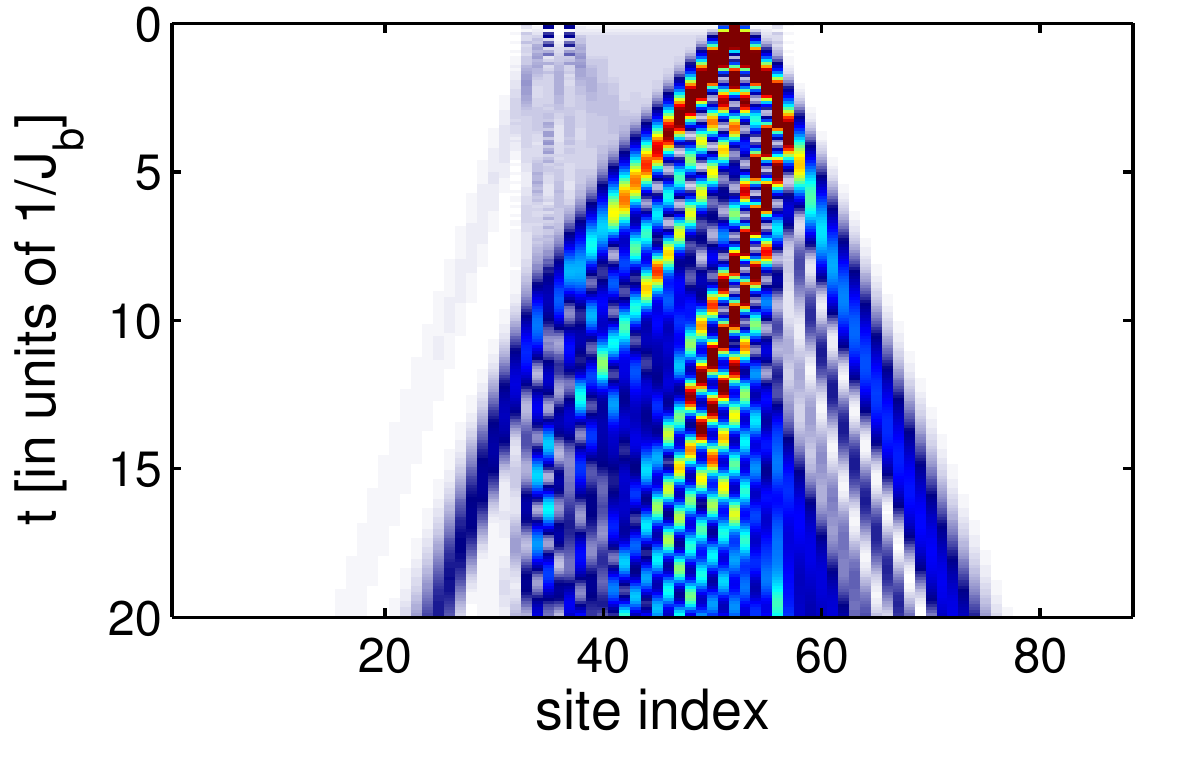}%
\caption{(Color online)
Dynamics of an initially localized unpaired particle $\mathrm{b}$, 
or $\mathrm{a}$-hole, in the cluster (left column), and an unpaired 
particle $\mathrm{a}$, $\mathrm{b}$-hole, in the cluster (right column),
for the initial cluster state 
$\tket{4_{-\mathrm{a}}}{2}{8}\tket{\cdot}{2}{8}\tket{4_{-\mathrm{b}}}{2}{8}$.
All parameters are as in Fig.~\ref{fig:two_image_mfq_fff},
and the bond dimension of the TEBD is $\chi=200$.}
\label{fig:two_image_lff_ffl}
\end{figure}

Fig. \ref{fig:two_image_lff_ffl} illustrates the results for 
a pair of initially localized defects of different type. Again, 
for $\Ja = \Jb$, both defects easily leave the cluster through 
its walls, but when $\Ja \neq \Jb$, only a fraction of 
the population of each defect leaves the cluster after the first
collision with its wall. Note, however, that since the two types 
of hole defects have different effective mass, their collisions 
with each other and the walls of the cluster can effectively 
redistribute their quasi-momenta, and no trimer defects are required 
to purify the MI cluster.

\section{Summary}

To conclude, in one-dimensional MI clusters of repulsively-bound dimers 
of bosons \cite{Petrosyan2007}, hole defects (unpaired particles, 
or monomers) can evaporate through the cluster boundaries, taking away 
the entropy of the system. In the case of dimers of identical bosons, 
only part of the monomer population can leave the cluster unassisted. 
Complete evaporation of the hole defects is possible in the presence 
of catalyzing particle defects (trimers), which efficiently thermalize
the hole defects via the quasi-momentum redistributing collisions. 
The particle defects themselves can not leave the cluster, due to 
the large energy mismatch $2U$ between a single excess particle on 
top of the $n=2$ MI cluster and on an empty lattice. 

In the case of dimers composed of two different bosonic species, 
the defect evaporation proceeds by itself, without the need of
any catalyzing species.

The system studied in this paper is amenable to experimental investigations
with cold atoms in optical lattices \cite{Bloch2008}. To prepare the cluster
of dimers surrounded by lattice vacuum, one starts with an optical lattice 
superimposed by a shallow confining potential populated by the MI phases with 
occupation numbers of $n=0,1,2$ in successive spatial shells \cite{Campbell2006, Folling2006}, 
followed by removal of all the atoms outside the central $n=2$ MI region \cite{Wuertz2009}. 
The homogeneous lattice potential is then achieved by turning off the shallow confining 
potential, while the spatial distribution of the defects and their dynamics inside and
outside the dimer cluster can be resolved using non-destructive single-site addressing 
techniques \cite{Bakr2009, Bakr2010, Sherson2010}. 

\acknowledgments

D.~M. and M.~F. acknowledge financial support through 
the SFB/TR49 of the Deutsche Forschungsgemeinschaft.
D.~P. acknowledges the support of the Humboldt Foundation.

\appendix

\section{Transmission of a particle through a domain wall}
\label{sec:apptrans}

Here we calculate the probability of transmission of a particle 
with quasi-momentum $k$ through a domain wall, as per Eq.~(\ref{eq:HTrans}). 
For the particle incident from the left, we solve the stationary 
Schr\"odinger equation using the standard scattering ansatz for 
the wave function,
\begin{equation}
 \psi_j = \left\{\begin{array}{lr}
                  e^{ikj}+\rho e^{-ikj}&j\le0\\
                  \tau e^{ik'j}&j\ge0
                 \end{array}\right. , \label{eq:statscatwf}
\end{equation}
where $\rho$ and $\tau$ are the complex reflection and transmission amplitudes. 
The energy eigenvalue is $E_k = E_k^{(\rm A)} = -2\JA\cos(k) = -2\JB\cos(k')
= E_{k'}^{(\rm B)}$, and therefore the refraction is given by
\begin{equation}
 k' = \cos^{-1}\left(\frac{\cos(k)}{\alpha}\right).
\end{equation}
Thus the transmission vanishes if 
$\cos(k) \ge \alpha$, where $\alpha = \JB/\JA$.
(In this paper we are primarily concerned with the case of $\alpha=1/2$,
except for Sec.~\ref{sec:twospecies}.)

Continuity at $j=0$ implies $1+\rho = \tau$, which together with 
the Schr\"odinger equation at $j=0$,
\begin{equation}
(\hH\psi)_0 = -\JA\psi_{-1} -\JB\psi_{1} = E_k \psi_{0},
\end{equation}
yields
\begin{equation}
 \rho = \frac
{  \JB e^{ik'}  + \JA e^{-ik} - 2\JA\cos(k)}
{ -\JB e^{ik'}  - \JA e^{ ik} + 2\JA\cos(k)} .
\end{equation}

The current density in the two parts of the system is given by
\begin{equation}
f_j = \left\{\begin{array}{lr}
              f_j^{({\rm A})} 
              = -i\JA (\psi_j^*\psi_{j+1}-\psi_j\psi_{j+1}^* ) & j < 0 \\
              f_j^{({\rm B})} 
              = -i\JB (\psi_j^*\psi_{j+1}-\psi_j\psi_{j+1}^* ) & j \ge 0
             \end{array}\right. .
\end{equation}
One can readily verify that
\begin{equation}
 \frac{\rm d}{{\rm d}t}\psi^*_j\psi_j 
= (-i\hH\psi )^*_j \psi_j + \psi_j^* (-i\hH\psi )_j = - (f_j-f_{j-1}).
\end{equation}
For the state of Eq.~(\ref{eq:statscatwf}), we have
\begin{subequations}
\begin{eqnarray}
 f_{\rm in} &=& 2\JA \sin(k),\\
 f_{\rm ref} &=& -2\JA \sin(k)\rho^*\rho, \\
 f_{\rm trans} &=& 2\JB \sin(k') \tau^*\tau,
\end{eqnarray}
\end{subequations}
so that the reflection and transmission probabilities are
\begin{eqnarray}
R &=& \left\vert\frac{f_{\rm ref}}{f_{\rm in}}\right\vert = \rho^*\rho , \\
T &=& \left\vert\frac{f_{\rm trans}}{f_{\rm in}}\right\vert 
= \alpha\frac{\sin(k')}{\sin(k)}\tau^*\tau . \label{eq:transm}
\end{eqnarray}
On can verify that $T+R=1$ as it should.

\section{Two particle Hamiltonian in quasi-momentum space}
\label{sec:appmomentum}

We consider a pair of distinguishable, locally interacting particles 
on a lattice described by Hamiltonian
\begin{eqnarray}
\label{eq:Hmom}
 \hH &=& -J_a \left( \sum_{j=1}^{L-1} \had_j\ha_{j+1} 
 + \gamma \had_L\ha_{1} +  \hadj \right)  
 \nonumber\\ & & 
 -J_t \left( \sum_{j=1}^{L-1} \htrd_j\htr_{j+1} 
 + \gamma \htrd_L\htr_{1} + \hadj \right)
 \nonumber \\ & & 
 + U \sum_{j=1}^L \had_j\ha_{j}\htrd_j\htr_{j} ,
\end{eqnarray}
where the periodic and open boundary conditions correspond, respectively,
to $\gamma = 1$ and $\gamma = 0$. The operators $\had_j$ and $\htrd_j$ 
create soft-core particles interacting via $U$, which is convenient 
for the exact numerical simulations presented in Sec.~\ref{sec:momrd:ed}.

In quasi-momentum representation, $k=2\pi\nu/L$ ($\nu = \lfloor-\frac{L}{2}+1\rfloor\ldots\lfloor\frac{L}{2}\rfloor$), 
we have $\had_k = \frac{1}{\sqrt{L}}\sum_{j=1}^{L} e^{i k j} \had_j$ and
$\htrd_k = \frac{1}{\sqrt{L}}\sum_{j=1}^{L}e^{i k j} \htrd_j$
and two such particles have a probability $L^{-1}$ to be on the same real lattice site. 
The Hamiltonian then reads
\begin{eqnarray}
& & \hH = \nonumber \\ & &
-2J_a\sum_{k} \cos (k) \had_k\ha_k  
  + \frac{(1-\gamma)J_a}{L} \sum_{k,k'} \had_k\ha_{k'} (e^{i k }+e^{-i k'}) 
  \nonumber \\ & & 
  -2J_t\sum_{k} \cos (k) \htrd_k\htr_k  
 + \frac{(1-\gamma)J_t}{L} \sum_{k,k'} \htrd_k\htr_{k'} 
 (e^{i k}+e^{-ik '}) 
 \nonumber \\ & & 
  + \frac{U}{L} \sum_{k,k',k''} \had_k\ha_{k'} \htrd_{k''}\htr_{(k+k''-k')} .
\end{eqnarray}

\section{Effective theory for monomers and trimers}
\label{sec:appeffective}

The non-locality of the effective theory presented in Sec.~\ref{sec:effdefHam}
might seem surprising at first sight. From the point of view of quantum 
information theory, however, the Hamiltonian (\ref{eq:Heff}) is still local,
in the sense that the commutator $[\tH_j, \tH_{j'}]$ vanishes except for 
$j' = j\pm1$, despite the fact that the support of any two 
$\tH_j, \tH_{j'}$ has a large overlap. This property should always be 
conserved in any effective theory, since it guarantees that correlations 
in the model system travel with the same maximal velocity as in the full system 
\cite{Bravyi2006, Eisert2006}. This property also permits the application 
of the TEBD numerical method, in conjunction with the conservation 
of the total particle number, to the effective model. For then all the 
basis states used in the TEBD (eigenstates of the reduced density matrices 
for all bi-partitions of the lattice) are, by construction, the eigenstates 
of the total particle number in the corresponding subsystem. Since the total
particle number is the only observable that enters Hamiltonian~(\ref{eq:Heff})
via $\hat{P}_j^{\nr}$, this type of non-locality does not introduce additional 
difficulties in the use of the TEBD method.

The effective model can also be extended to higher orders in perturbation 
theory. In second order, this introduces nearest neighbor interactions,
local potentials and effective exchange between monomers and trimers. 
All these terms are of the order of $J^2/U$ and depend on $\Theta$, 
which can now assume four different values depending on the type of 
bond between sites $j$ and $j+1$. Another term of the same order describes 
hole defect hopping to the next-nearest neighbor site in the cluster. 
As this is spanning three sites, it also depends on the state of the 
central site and requites more values of $\Theta$. The presence of 
such a longer-range term would necessitate a more general numerical 
simulation algorithm than TEBD. We have verified, however, that the
effective Hamiltonian (\ref{eq:extHubtrim}) containing only the terms 
first order in $J$ already captures all the essential physics discussed 
in this paper.

\begin{figure}[t]
\centering
(a)\includegraphics[width=.45\columnwidth]{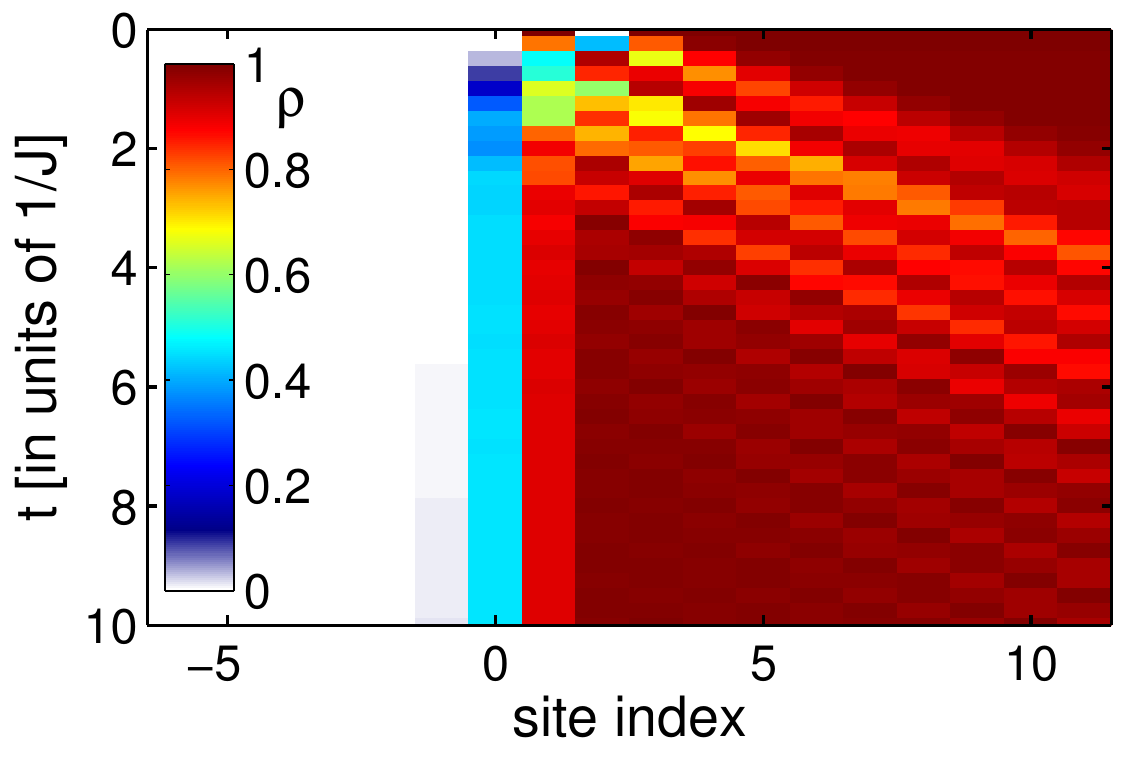}
(b)\includegraphics[width=.45\columnwidth]{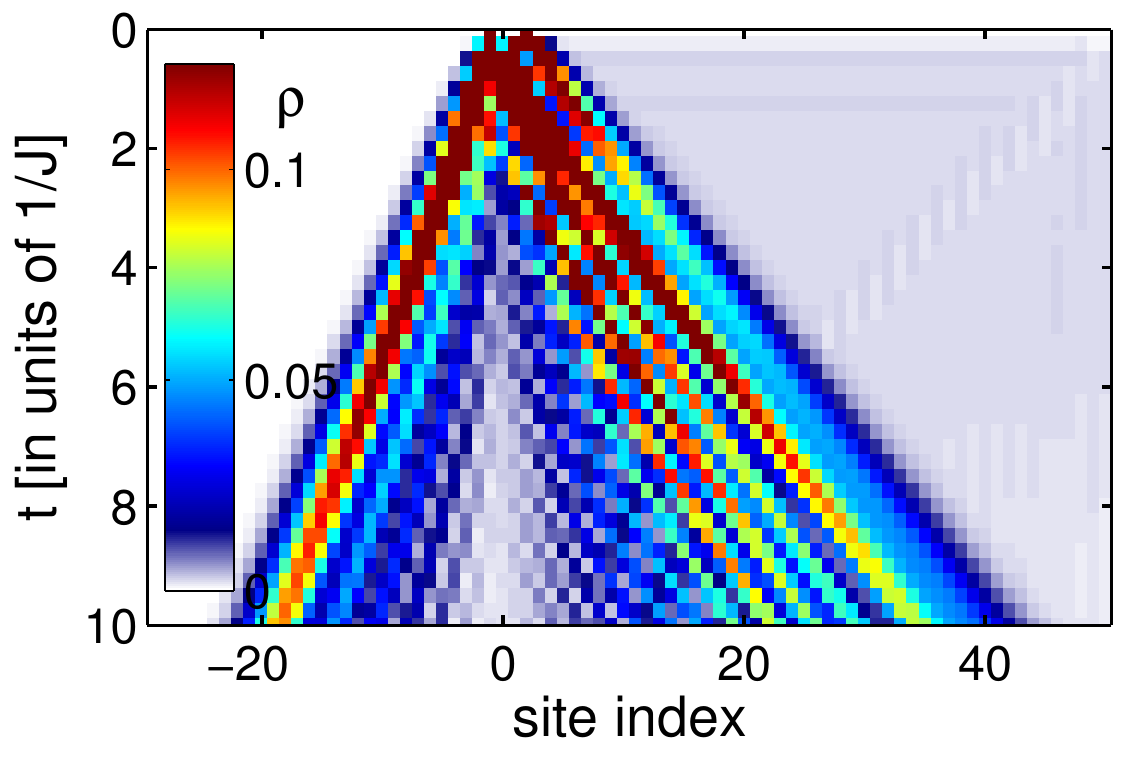}\\
(c)\includegraphics[width=.45\columnwidth]{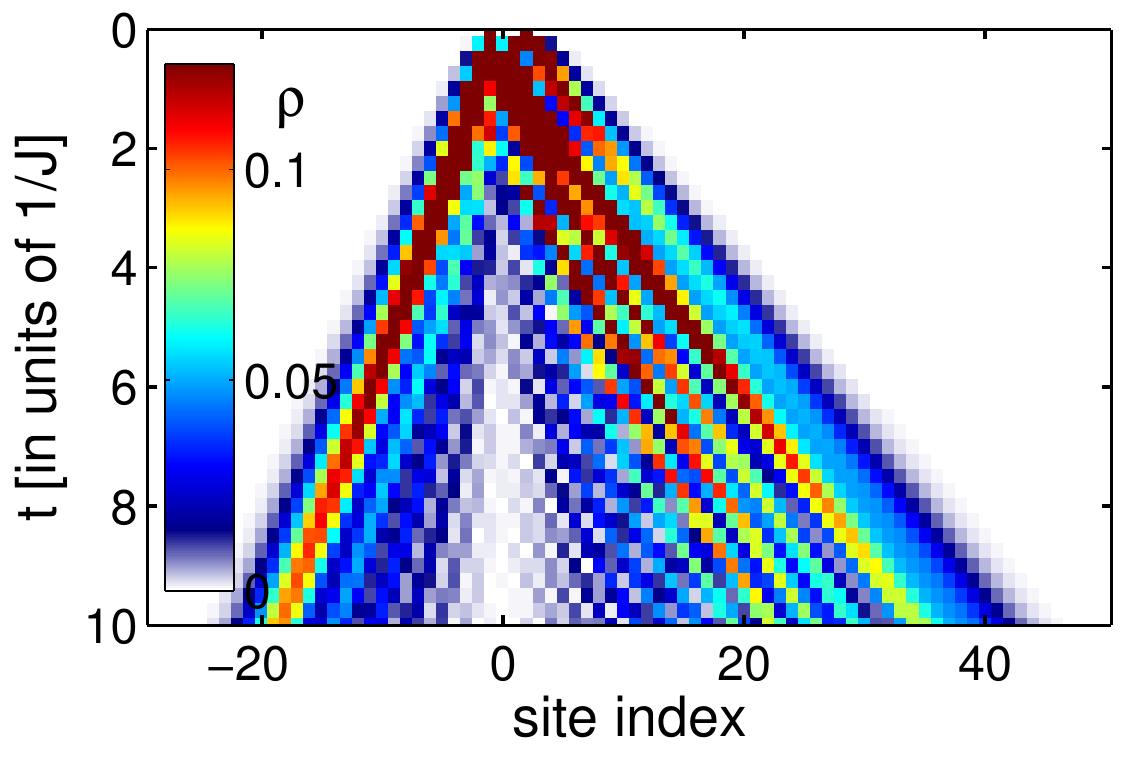}
(d)\includegraphics[width=.45\columnwidth]{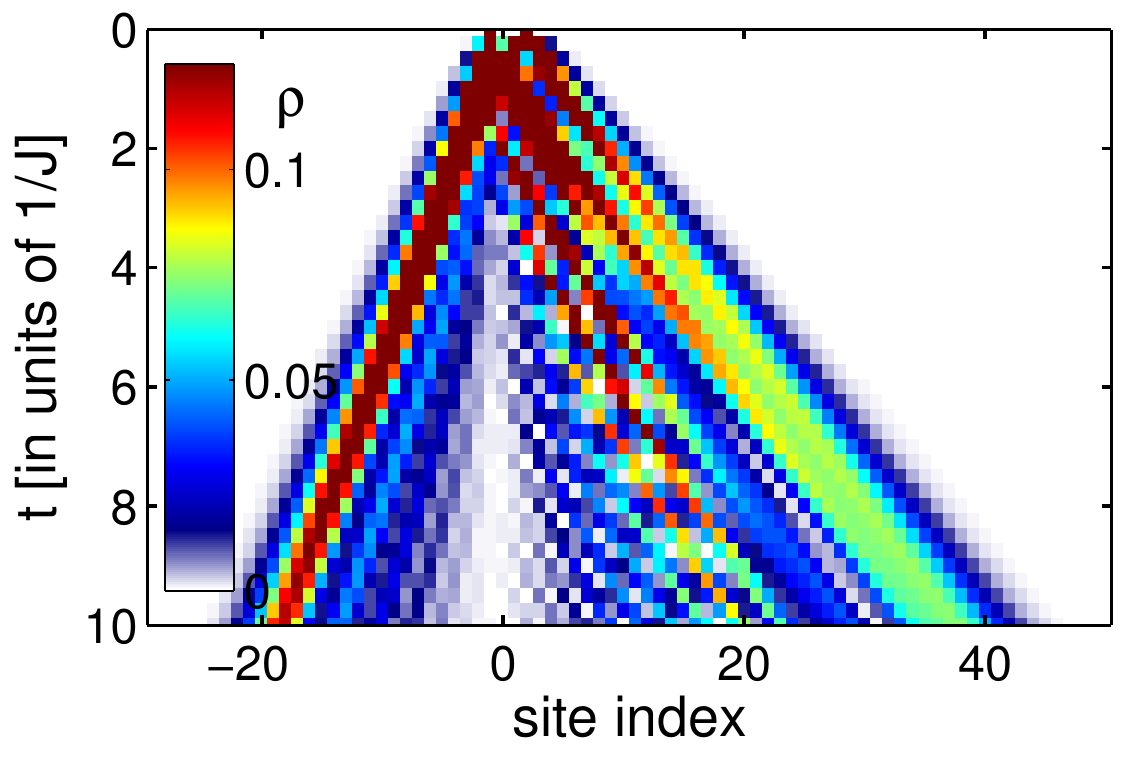}
\caption{(Color online) 
Comparison of the full Bose-Hubbard dynamics, Eq.~(\ref{eq:HBH}) with $U=100J$, 
(a) and (b), with the effective model, Eq.~(\ref{eq:Heff}), (c). 
The initial state contains a MI cluster of dimers on sites $j \geq 1$ and 
localized monomers at sites $j=-1$ and $j=2$. 
The density of dimers is shown in (a), and the density of monomers in (b)-(d). 
Note that the cluster boundary is shifted upon particle crossing, 
which manifests in (a) as a smoothing of step in the dimer density. 
The effective Hamiltonian without moving boundaries, 
$\hH = \sum_{j=1}^{L-1} \hH_j^{[\Theta(j)]}$, yields the dynamics of (d).}
\label{fig:test2}
\end{figure}

In Fig.~\ref{fig:test2} we compare the dynamics of hole defects 
obtained from the full and effective models, which agree very well 
for large interaction strength $U \gg J$. 
Observe, however, that a local theory neglecting the motion of 
the cluster boundaries, Fig.~\ref{fig:test2}(d), and therefore 
violating the conservation of the total number of dimers and bare 
particles, does not describe the dynamics quantitatively correctly.

\section{MPS representation of the initial state}
\label{sec:appconstr}

Here we show how to construct an exact MPS representation for a lattice 
containing fixed number of bosons each in a certain single particle 
eigenstate. The resulting MPS will be in the canonical representation 
\cite{Perez-Garcia2007} and symmetric \cite{Schollwock2011}, 
i.e., it will be an eigenstate of the total particle 
number by construction. The construction is analogous to that of matrix 
product operators for fixed total particle number \cite{Muth2011a}.

The single particle state is given by a normalized wave-function 
$\phi_j$. In the examples of Sec. \ref{sec:MBinst}, we have 
$\phi_j = \frac1{\sqrt{L}}e^{ikj}$ with fixed quasi-momentum $k$. 
The corresponding state 
$\ket{\Psi^1} = \sum_{j=1}^L \phi_j \had_j \ketv$ can be written as
\begin{equation}
 \ket{\Psi^1} = \big( \sqrt{\qA} \hadA + \sqrt{\qB} \hadB \big) 
\ketvA \otimes \ketvB ,
\end{equation}
where sub-lattice A spans sites $1$ to $m$ and sub-lattice B 
is from $m+1$ to $L$, while $\qA=\sum_{j=1}^m \phi_j^*\phi_j$ is 
the single particle probability of being in A.
The bosonic creation operators $\hadA$ and $\hadB$ are defined by
\begin{equation}
 \hadA = \frac{1}{\sqrt{\qA}} \sum_{j=1}^m \phi_j\had_j, \quad
 \hadB = \frac{1}{\sqrt{\qB}} \sum_{j=m+1}^L \phi_j\had_j . 
\end{equation}

The state of the lattice with $N$ particles in the same 
single particle state can then be expressed as
\begin{eqnarray}
 \ket{\Psi^N} &=& \frac{1}{\sqrt{N!}} \big( \sqrt{\qA} \hadA 
 + \sqrt{\qB}\hadB \big)^N \ketvA \otimes \ketvB 
\nonumber \\
&=& \frac{1}{\sqrt{N!}} \sum_{l=0}^N \binom{N}{l}  
\\ &&\times
\big( \sqrt{\qA}\hadA \big)^l 
\big( \sqrt{\qB}\hadB \big)^{N-l} \ketvA\otimes\ketvB , 
\nonumber
\end{eqnarray}
and the density matrix of the system is
\begin{eqnarray}
& & \ket{\Psi^N}\bra{\Psi^N} 
\nonumber \\ & & 
= \frac{1}{N!} \sum_{l,l'=0}^N 
\binom{N}{l} \binom{N}{l'} 
\nonumber \\ & &\times 
\big( \sqrt{\qA}\hadA \big)^l \ketvA\bravA \big( \sqrt{\qA}\haA \big)^{l'} 
\nonumber \\ & &\otimes 
\big( \sqrt{\qB}\hadB \big)^{N-l} \ketvB \bravB 
\big( \sqrt{\qB}\haB \big)^{N-l'} . 
\end{eqnarray}
The density matrix of subsystem A is
\begin{eqnarray}
 \rho_{\rm A} &=& \Tr_{\rm B}\big[\ket{\Psi^N}\bra{\Psi^N} \big] 
\nonumber \\ 
&=& \frac{1}{N!} \sum_{l=0}^N \binom{N}{l}^2 (N-l)! \, (\qB)^{N-l} 
\nonumber \\ &&\times 
\big( \sqrt{\qA}\hadA \big)^l \ketvA\bravA \big( \sqrt{\qA}\haA \big)^{l}.
\end{eqnarray}
Note that $\rho_{\rm A}$ has at most $\chi = N+1$ nonzero eigenvalues, 
one for each possible distribution of the $N$ particles between A and B. 
With $\hat{P}^{[\rm A]}_l$ the projector onto the $l$ particle sector 
of subsystem A, the probability of finding $l$ particles in A is
\begin{eqnarray}
 \Tr_{\rm A}\big[\rho_{\rm A}\hat{P}^{[\rm A]}_l\big] &=& 
\frac{1}{N!} \sum_{l=0}^N \binom{N}{l}^2 (N-l)!\ l!\ (\qB)^{N-l} (\qA)^l 
\nonumber\\
 &=& \sum_{l=0}^N \binom{N}{l} (\qB)^{N-l} (\qA)^l 
\nonumber\\
 &=& B_{\qA}(l|N),
\end{eqnarray}
which is a binomial distribution.

We can now construct $\ket{\Psi^N}$ as a matrix product state in the 
canonical \cite{Perez-Garcia2007} form. Given a bi-partition of the lattice, 
its Schmidt decomposition is
\begin{equation}
 \ket{\Psi^N} = \sum_{l=0}^N \lambda^{[m]}_l \ket{\Psi^l}_{\textrm A} 
\otimes \ket{\Psi^{N-l}}_{\textrm B},
\end{equation}
with $\ket{\Psi^l}_{\textrm A} = \frac{1}{\sqrt{l!}} (\hadA)^l\ketvA$ and 
$\ket{\Psi^{N-l}}_{\textrm B} = \frac{1}{\sqrt{(N-l)!}} (\hadB)^{(N-l)}\ketvB$.
The MPS will have bond dimension of $\chi=N+1$. 
The probability of finding $l$ particles to the left of bond $m$ is
\begin{equation}
\big( \lambda^{[m]}_l \big)^2 = B_{\qA}(l|N) . \label{eq:lambda}
\end{equation}
We then continue with the Schmidt decomposition at the following bond. 
The remaining task is to determine the coefficients of
\begin{eqnarray}
 \ket{\Psi^N} &=& \sum_{l=0}^N \sum_{r=l}^N
 \lambda^{[m]}_l \Gamma^{[m+1]}_{lr} \lambda^{[m+1]}_r 
\nonumber \\ & & \times 
\ket{\Psi^l}_{\textrm A} \otimes  \ket{\Psi^{r-l}}_{m+1} 
\otimes \ket{\Psi^{N-r}}_{\textrm B'}.
\end{eqnarray}
The $\lambda$ tensors are already known from (\ref{eq:lambda}). 
The sub-chain ${\rm B}'$ comprises sites $m+2$ to $L$. Thus 
$\big(\lambda^{[m]}_l \big) ^2 \big|\Gamma^{[m+1]}_{lr} \big|^2 
\big( \lambda^{[m+1]}_r \big)^2$ is the probability of finding 
$N-r$ particles on the right of bond $m+1$ \emph{and} $l$ 
particles on the left of bond $m$, resulting in
\begin{eqnarray}
\big| \Gamma^{[m+1]}_{lr} \big|^2 &=& 
\frac{B_{\frac{q_m}{q_{m+1}}}(l|r)}{B_{q_m}(l|N)} \label{eq:gamma} \\
&=& \frac{r! \, (N-l)!}{(r-l)! \, N!}q_{m+1}^{-r} 
\big( q_{m+1} - q_{m} \big)^{r-l} \big( 1 - q_{m} \big)^{l-N} . \nonumber
\end{eqnarray}
For the phase to be correct, we obviously have to set
\begin{equation}
\arg \big(\Gamma^{[m+1]}_{lr} \big) = (r-l) \,  \arg \big( \phi_{m+1} \big).
\label{eq:gammaphase}
\end{equation}
Equations (\ref{eq:lambda}), (\ref{eq:gamma}), and (\ref{eq:gammaphase}) 
completely determine the tensors $\Gamma$ and $\lambda$.
Note that in this particular case, the value of the bond index of 
$\lambda^{[m]}$ has a physical meaning of the number of particles 
to the left of bond $m$. The resulting MPS is an eigenstate 
of the total particle number, which can be used in TEBD implementations 
that take advantage of particle number conservation explicitly, 
as in this paper.

The construction is more complicated when one intends to prepare $N_\alpha$ 
particles in different single particle states $\alpha=1,2,\dots, M$.
From simple combinatorial considerations, we deduce that the Schmidt rank 
will be $\chi = \prod_\alpha(1+N_\alpha)$, i.e., exponentially large 
in the number $M$ of different single particle states. (This implies that, 
as a starting point for dynamical simulations, one can construct 
an exact MPS for the ground state of non-interacting bosons, as done 
in \cite{Muth2010b}, but not for non-interacting fermions.)
The exact expression in terms of the $q_{m,\alpha}$ will contain 
overlaps between the different single particle states, which in general 
are finite in any subsystem even if the single particle states are 
orthogonal on the entire lattice.


\end{document}